\documentclass[aps,prb,twocolumn,superscriptaddress,showpacs]{revtex4}
\usepackage{graphicx}
\usepackage[T1]{fontenc}
\usepackage{mathrsfs}
\usepackage{amssymb}
\usepackage{graphicx}
\usepackage{latexsym}
\usepackage{amsmath}
\usepackage{amsfonts}
\usepackage{bm}
\usepackage{multirow}
\usepackage{color}
\usepackage{comment}

\newcommand{\beq}{\begin{equation}}
\newcommand{\eeq}{\end{equation}}
\newcommand{\beqn}{\begin{eqnarray}}
\newcommand{\eeqn}{\end{eqnarray}}

\DeclareMathAlphabet{\mathbbold}{U}{bbold}{m}{n}

\makeatletter
\newcommand\xleftrightarrow[2][]{%
\ext@arrow 9999{\longleftrightarrowfill@}{#1}{#2}}
\newcommand\longleftrightarrowfill@{%
\arrowfill@\leftarrow\relbar\rightarrow} \makeatother

\begin{document}

\title{Lattice models for Non-Fermi Liquids with Tunable Transport Scalings}

\author{Xiao-Chuan Wu}
\affiliation{Department of Physics, University of California,
Santa Barbara, CA 93106, USA}

\author{Chao-Ming Jian}
\affiliation{Kavli Institute of
Theoretical Physics, Santa Barbara, CA 93106, USA}
\affiliation{ Station Q, Microsoft Research, Santa Barbara,
California 93106-6105, USA}

\author{Cenke Xu}
\affiliation{Department of Physics, University of California,
Santa Barbara, CA 93106, USA}

\date{\today}
\begin{abstract}

A variety of exotic non-fermi liquid (NFL) states have been
observed in many condensed matter systems, with different scaling
relations between transport coefficients and temperature. The
``standard'' approach to studying these NFLs is by coupling a
Fermi liquid to quantum critical fluctuations, which potentially
can drive the system into a NFL. In this work we seek for an
alternative understanding of these various NFLs in a unified
framework. We first construct two ``elementary'' {\it
randomness-free} models with {\it four-fermion} interactions only,
whose many properties can be analyzed exactly in certain limit
just like the Sachdev-Ye-Kitaev (SYK) model. The most important
new feature of our models is that, the fermion scaling dimension
in the conformal invariant solution in the infrared limit is
tunable by charge density. Then based on these elementary models,
we propose two versions of lattice models with four fermion
interactions which give us non-fermi liquid behaviors with DC
resistivity scaling $\varrho \sim T^\alpha$ in a finite
temperature window, and $\alpha \in [1, 2)$ depends on the fermion
density in the model, which is a rather universal feature observed
in many experimental systems.

\end{abstract}

\maketitle

\section{Introduction}

Non-fermi liquid (NFL) states represent a family of exotic
metallic states that do not have long-lived quasi-particles, and
hence behave fundamentally differently from the standard Landau
Fermi liquid
theory~\cite{hertz,millis,nfl1,polchinskinfl,nayaknfl1,nayaknfl2,nematicnfl,nfl2,nfl3,nfl4,nfl5,nfl6,nflqmc}.
The most well-known NFL, the ``strange metal'' phase at the
optimal doping of the cuprate high temperature superconductors,
has a universal scaling of its DC resistivity $\varrho \sim
T$~\cite{linear1,linear2,linear3,linear4,Varma1989}, while the
standard Fermi liquid theory predicts $\varrho \sim T^2$. Recently
the same strange metal behavior was observed in twisted bilayer
graphene above the superconductor phase~\cite{pablostrange}. A
consensus of the nature of the strange metal phase has not been
reached yet, but a series of toy models, despite their relatively
unnatural forms, seem to capture many of the key universal
features of the strange metal phase. These models are the
so-called Sachdev-Ye-Kitaev (SYK) model and its
generalizations~\cite{SachdevYe1993,Kitaev2015,Sachdev2015,Polchinski2016,MaldacenaStanford2016,Witten2016,Klebanov2016,Gross2017}.
For example, it was found that the SYK model has marginally
relevant ``pairing instability'' just like the ordinary Fermi
liquid state~\cite{xu2017,pairex}, which is consistent with the
fact that the non-Fermi liquid phase is often preempted by a dome
of ``ordered phase'' with pair condensate of fermions
(superconductivity) at low
temperature~\cite{nflpairing1,nflpairing2,nflpairing3,nflpairing4,nflpairing5,nflpairing6,inter}.
Thus the ``SYK phase'' can be viewed as a candidate parent phase
of superconductor. Also, the recently observed anomalous charge
density fluctuation of the strange metal~\cite{abbamonte} suggests
connection to the SYK model~\cite{SachdevYe1993}.
Last but not least, a series of
generalizations based on the SYK model has shown linear-$T$
resistivity for a large temperature window, and the scaling
dimension of the fermion operators in the SYK model is the key for
the linear-$T$ scaling of the
resistivity~\cite{Song2017,Gu2017b,patel2017,berg2018}. But these
models, in order to ensure solubility, require fully random
four-fermion interactions with a Gaussian distribution and zero
mean, which is unlikely to exist in real materials. More recently
a model on the square lattice without random interaction was
constructed~\cite{xu2018}, which in the soluble limit mimics the
physics of the so called three-index tensor
models~\cite{Witten2016,Klebanov2016,Gurau}, and gives us the same
desirable physics such as linear$-T$ scaling of DC resistivity,
and marginally relevant instability towards superconductor and
other competing phases.

Most of the previously discussed generalizations of the SYK model
aimed at constructing the strange metal phase with precisely
linear$-T$ scaling of resistivity. But NFL can have much richer
physics than the strange metal. In various systems with NFL
behaviors, the DC resistivity can scale with temperature as
$\varrho \sim T^\alpha$ with $1 \leq \alpha <
2$~\cite{rho0,cecusi,rho1,rho2,rho3,review}, and $\alpha$ is
usually tunable by varying the charge density. As we mentioned in
the previous paragraph, the linear$-T$ scaling of the DC
resistivity is a direct consequence of the scaling dimension
$\Delta_f = 1/4$ of the fermion operator in the SYK model after
disorder average. To design a model with $\alpha$ between 1 and 2,
we can in principle start with the SYK$_q$ model with $q
> 4$. But these models require a $q-$body interactions between the
fermions, and hence are also not realistic for condensed matter
systems. Thus to construct a relatively realistic NFL with
$\varrho \sim T^\alpha$ and an arbitrary $\alpha \in [1,2)$, we
need to start with a model with {\it four-fermion interaction
only} and {\it no randomness}, but with conformal solutions whose
fermion scaling dimensions can be different from $1/4$. And most
ideally the fermion scaling dimension is tunable with charge
density.


The standard approach of understanding these NFLs is by coupling
the Fermi liquid state to a fluctuating bosonic quantum critical
mode, and the relevant boson-fermion coupling can potentially
drive the system into a
NFL~\cite{hertz,millis,nfl1,polchinskinfl,nayaknfl1,nayaknfl2,nematicnfl,nfl2,nfl3,nfl4,nfl5,nfl6,nflqmc}.
And the transport-temperature scaling would depend on the spatial
dimensionality and also the momentum carried by the quantum
critical mode. In this paper we take a different approach. We will
first design two elementary models for interacting fermions that
is free of randomness, whose solution in certain theoretical limit
is a conformal field theory, and most importantly the fermion has
a scaling dimension that depends on the charge density of the
model. Then based on these elementary models we design two
versions of lattice models which naturally give us $\varrho \sim
T^\alpha$, and $\alpha \in [1,2)$ is tunable by charge density.
Our models provide an alternative approach of studying various
experimentally observed NFLs in a unified framework.

\section{The elementary models}

\begin{figure}
\includegraphics[scale=0.2]{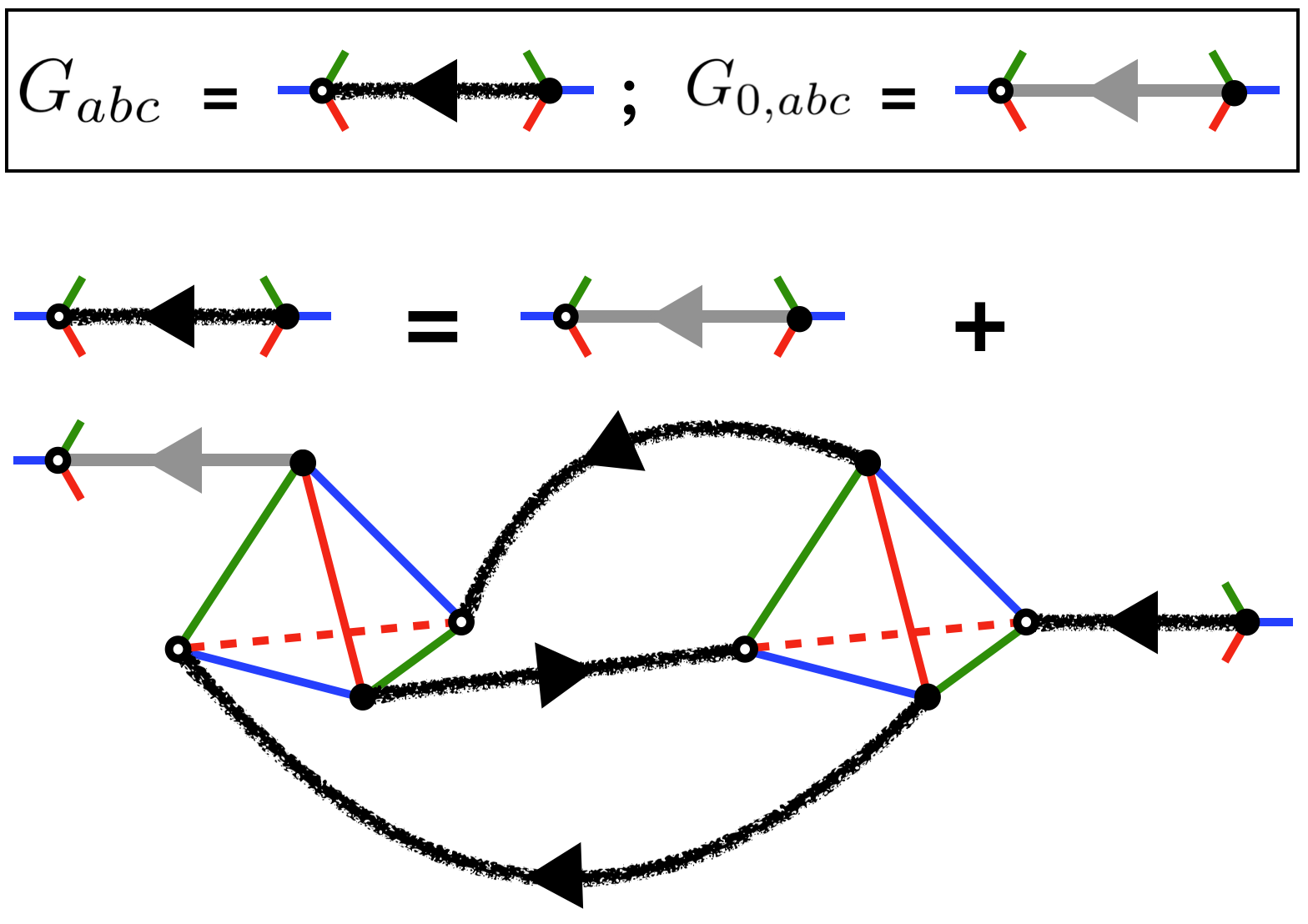}
\caption{The large-$N$ Schwinger-Dyson equation for various complex tetrahedron models. }
\label{melon}
\end{figure}

We first give a brief review of the ``tetrahedron'' three-index
tensor model without any disorder, and in the large-$N$ limit
their solutions mimic the better-known $\textrm{SYK}_{4}$ model.
As was discussed in Ref.~\onlinecite{Klebanov2016}, the original
$\textrm{U}\left(N_{a}\right)\times\textrm{U}\left(N_{b}\right)\times\textrm{O}\left(N_{c}\right)$
symmetric tetrahedron model can be written as
\begin{equation}
H=\frac{g}{\sqrt{N_{a}N_{b}N_{c}}}\psi_{a_{1}b_{1}c_{1}}^{\dagger}\psi_{a_{2}b_{2}c_{1}}^{\dagger}\psi_{a_{1}b_{2}c_{2}}\psi_{a_{2}b_{1}c_{2}},
\end{equation}
where $a=1,\ldots,N_{a}$, $b=1,\ldots,N_{b}$, $c=1,\ldots,N_{c}$.
One can prove that as long as
\begin{equation}
0<\frac{N_{a}}{N_{b}},\frac{N_{b}}{N_{c}},\frac{N_{c}}{N_{a}}<\infty
,
\end{equation}
this tensor model is dominated by the melonic diagrams in the
large-$N_a, N_b, N_c$ limit (Fig.~\ref{melon}), and its solution
is a conformal field theory fixed point in the infrared limit. At
the conformal fixed point, the melonic diagrams can be summed by
solving the Schwinger-Dyson equations which are identical to the
original SYK$_4$ model for the complex
fermions~\cite{SachdevYe1993,Sachdev2015,Kitaev2015}:
\begin{flalign}
 G\left(i\omega_{n}\right)&=\frac{1}{i\omega_{n}+\mu-\Sigma\left(i\omega_{n}\right)},\\\Sigma\left(\tau\right)&=-4g^{2}G\left(\tau\right)^{2}G\left(-\tau\right),
\end{flalign}
where the two-point Green's function $G\left(\tau\right)$ is
defined as
\begin{equation}
 G\left(\tau\right)\delta_{aa^{\prime}}\delta_{bb^{\prime}}\delta_{cc^{\prime}}=-\left\langle \mathbb{T}_{\tau}\psi_{abc}\left(\tau\right)\psi_{a^{\prime}b^{\prime}c^{\prime}}^{\dagger}\left(0\right)\right\rangle
\end{equation}
$\Sigma$ is the self energy, $\omega_{n}$ is fermionic Matsubara
frequency $\omega_{n}=\left(2n+1\right)\pi T,n\in\mathbb{Z}$, and
$\tau$ is imaginary time. One key feature of this model is that in
its conformal solution the fermions have the scaling dimension
\begin{equation}
\Delta_{\psi}=\frac{1}{4} \label{scaling 1/4}
\end{equation}
just like the SYK$_4$ model.

This model certainly has many variants with the same large-$N$
solution. In Ref.~\onlinecite{xu2018} in order to make connection
to the cuprates, we constructed a lattice model based on a
modified tensor model with the form
 \begin{equation}
 H=\frac{g\mathcal{J}_{c_{1}c_{1}^{\prime}}\mathcal{J}_{c_{2}c_{2}^{\prime}}}{\sqrt{N_{a}N_{b}N_{c}}}
 \psi_{a_{1}b_{1}c_{1}}^{\dagger}\psi_{a_{2}b_{2}c_{1}^{\prime}}^{\dagger}\psi_{a_{1}b_{2}c_{2}}\psi_{a_{2}b_{1}c_{2}^{\prime}},
 \end{equation}
where $\mathcal{J}$ is the antisymmetric matrix associated with
the $\text{Sp}\left(N_{c}\right)$ group and
$\mathcal{J}_{cc^{\prime}}\psi_{c}\psi_{c^{\prime}}$ forms an
$\text{Sp}\left(N_{c}\right)$ singlet.

So far all the tetrahedron models are comprised of one-orbital of
fermions with three indices and conformal dimension $1/4$ in the
soluble limit. In this paper, we consider generalizations to two
versions of ``elementary'' models each with two orbitals (types)
of fermions $\psi$ and $\chi$, and a mutual four-fermion
interaction. The existence of multi-orbitals of fermions is
analogous to the situation in many heavy fermion systems, where
most of the NFLs were observed. This simple generalization leads
to some important new features: the conformal dimensions
$\Delta_{\chi}$ and $\Delta_{\psi}$ can be tuned by changing the
parameters, especially the particle density in the models. These
elementary models enable us to build several lattice models for
NFLs with different transport scalings with randomness-free
four-fermion interactions.

\subsection{Model A}


The first ``elementary model'' we construct takes the following
form: \beqn H_{0}^A &=& \sum_{a_1, a_2, b_1, b_2 = 1}^N
\sum_{c=1}^{M_1} \sum_{d=1}^{M_2} \frac{g}{N\sqrt{M}} \cr\cr &&
\left( \psi^\dagger_{a_1, b_1, c} \psi_{a_2, b_2, c}
\chi^\dagger_{a_1, b_2, d} \chi_{a_2, b_1, d} + h.c. \right),
\label{H0A} \eeqn where $M = \sqrt{M_1 M_2}$. $\psi$ and $\chi$
are two orbitals (types) of fermions each carries three indices.
The model above is the simplest model with the desired features.
It has continuous symmetries just like the original tetrahedron
model, but these symmetries are not essential to our results.
There are also some discrete symmetries that are more important
for the solution, which will be spelled out later.

In the large-$N, M_1, M_2$ limit, just like the three-index tensor
models, only the ``melonic diagrams'' dominate. The sum of all the
melonic diagrams must satisfy the coupled Schwinger-Dyson (S-D)
equations:
 \begin{flalign}
G_{\psi}\left(i\omega_{n}\right)&=\frac{1}{i\omega_{n}+\mu_{\psi}-\Sigma_{\psi}\left(i\omega\right)},\\G_{\chi}\left(i\omega_{n}\right)&=\frac{1}{i\omega_{n}+\mu_{\chi}-\Sigma_{\chi}\left(i\omega\right)},
 \end{flalign}
and the self energies are
 \begin{flalign}
 \Sigma_{\psi}^{A}\left(\tau\right) & =-4g^{2}\sqrt{\frac{M_{2}}{M_{1}}}G_{\psi}\left(\tau\right)G_{\chi}\left(\tau\right)G_{\chi}\left(-\tau\right),
 \\
 \Sigma_{\chi}^{A}\left(\tau\right) &
 =-4g^{2}\sqrt{\frac{M_{1}}{M_{2}}}G_{\chi}\left(\tau\right)G_{\psi}\left(\tau\right)G_{\psi}\left(-\tau\right),
 \end{flalign}
where we have introduced different chemical potentials
$\mu_{\psi}, \mu_{\chi}$ for the two fermions to fix the particle
densities.

Apparently, in this model the particle density of $\psi$ and
$\chi$ are separately conserved, thus we can introduce filling
factor $\mathcal{Q}_{\psi}, \mathcal{Q}_{\chi} \in (0,1)$
separately. $\mathcal{Q}_{\psi}$ is defined as \beqn
\mathcal{Q}_{\psi} = \frac{\sum_{a,b,c} \langle
\psi^\dagger_{a,b,c}\psi_{a,b,c} \rangle}{N^2 M_1},
\label{ff}\eeqn and $\mathcal{Q}_{\chi}$ is defined accordingly.
The role of the filling factors will be specified later and
derived in detail in the supplementary material. With fixed
filling factors $\mathcal{Q}_{\psi}$ and $\mathcal{Q}_{\chi}$,
just like the original S-Y model Ref.~\onlinecite{SachdevYe1993},
we should set $\Sigma\left(i\omega_{n}=0\right)=\mu$. Thus, we can
redefine the self energy as
 \begin{equation}
 \tilde{\Sigma}_{\psi/\chi}\left(i\omega_{n}\right)=\Sigma_{\psi/\chi}\left(i\omega_{n}\right)-\mu
 \end{equation}
Now in the infrared limit, assuming the self-energy always
dominates the $i\omega_n$ term in the infrared, the S-D equations
are simplified as
 \begin{equation}
 G_{\psi}\left(i\omega_{n}\right)\tilde{\Sigma}_{\psi}\left(i\omega_{n}\right)=G_{\chi}\left(i\omega_{n}\right)\tilde{\Sigma}_{\chi}\left(i\omega_{n}\right)=-1. \label{S-D eq}
 \end{equation}

At general filling factors $\mathcal{Q}_{\psi}$ and
$\mathcal{Q}_{\chi}$, and at zero temperature $T=0$, we use the
following power law ansatz at complex frequency $z$
($\textrm{Im}\left(z\right)>0,\left|z\right|\ll g$) to solve the
S-D equations
 \begin{flalign}
 G_{\psi}\left(z\right) &
 =C_{\psi}\frac{e^{-i\left(\pi\Delta_{\psi}+\theta_{\psi}\right)}}{z^{1-2\Delta_{\psi}}},
 \label{ansatz1} \\
 G_{\chi}\left(z\right) &
 =C_{\chi}\frac{e^{-i\left(\pi\Delta_{\chi}+\theta_{\chi}\right)}}{z^{1-2\Delta_{\chi}}},
 \label{ansatz2}
 \end{flalign}
where the real parameters $C,\theta,\Delta$ satisfy
\begin{flalign}
 & C_{\psi}>0,\qquad-\pi\Delta_{\psi}<\theta_{\psi}<\pi\Delta_{\psi},\\
 & C_{\chi}>0,\qquad-\pi\Delta_{\chi}<\theta_{\chi}<\pi\Delta_{\chi}.
\end{flalign}

There are in general six unknowns that we need to solve for:
$C_{\psi/\chi}$, $\Delta_{\psi/\chi}$ and $\theta_{\psi/\chi}$.
But through the S-D equations which are exact in the large-$N,
M_1, M_2$ limit, we will be able to determine five of them:
$C_{\psi}^2 C_{\chi}^2$, $\Delta_{\psi/\chi}$ and
$\theta_{\psi/\chi}$. The scaling dimensions $\Delta_{\psi/\chi}$
are the most important quantities which will determine the scaling
of the transport coefficients, as we will calculate explicitly
later. In the large-$N, M_1, M_2$ limit, only the product
$C_{\psi}^2 C_{\chi}^2$ is determined, while $C_{\psi}$ and
$C_{\chi}$ may be determined separately through subleading
diagrams.

The S-D equation, or the melonic diagrams, demand that the self
energies at complex frequency $z,\textrm{Im}\left(z\right)>0$ take
the following form:
 \begin{flalign}
 \tilde{\Sigma}_{\psi}^{A}\left(z\right)&\propto C_{\psi}C_{\chi}^{2}\sqrt{\frac{M_{2}}{M_{1}}}
 e^{i\left(\pi\Delta_{\psi}+\theta_{\psi}\right)}z^{1-2\Delta_{\psi}}, \label{proportional1}
\\
 \tilde{\Sigma}_{\chi}^{A}\left(z\right)&\propto C_{\chi}C_{\psi}^{2}\sqrt{\frac{M_{1}}{M_{2}}}
 e^{i\left(\pi\Delta_{\chi}+\theta_{\chi}\right)}z^{1-2\Delta_{\chi}}. \label{proportional2}
 \end{flalign}
Eventually the coupled S-D equations Eq.~\ref{S-D eq} lead to the
following self-consistent equations:
 \begin{flalign}
 2g^{2}C_{\psi}^{2}C_{\chi}^{2}\sqrt{\frac{M_{2}}{M_{1}}}\frac{\cos\left(2\pi\Delta_{\psi}\right)+\cos\left(2\theta_{\chi}\right)}{\pi\left(1-2\Delta_{\psi}\right)\sin\left(2\pi\Delta_{\psi}\right)} & =1, \label{self-consistent eq1}
 \\
 2g^{2}C_{\chi}^{2}C_{\psi}^{2}\sqrt{\frac{M_{1}}{M_{2}}}\frac{\cos\left(2\pi\Delta_{\chi}\right)+\cos\left(2\theta_{\psi}\right)}{\pi\left(1-2\Delta_{\chi}\right)\sin\left(2\pi\Delta_{\chi}\right)}  \label{self-consistent eq2}
 & =1.
 \end{flalign}
The conformal dimensions $\Delta_\psi$ and $\Delta_{\chi}$ also
must satisfy another relation, which physically guarantee that the
system is at a fixed point controlled by the four fermion
interaction: \beqn 2\Delta_{\psi}+2\Delta_{\chi}=1. \label{Dcons}
\eeqn

Additionally, the filling factors $\mathcal{Q}_{\psi}$ and
$\mathcal{Q}_{\chi}$ give further constraints on
$\Delta_{\psi/\chi}$, and $\theta_{\psi/\chi}$ (please refer to
the supplementary material):
\begin{flalign}
\mathcal{Q}_{\psi} &
=\frac{1}{2}-\frac{\theta_{\psi}}{\pi}-\left(\frac{1}{2}-\Delta_{\psi}\right)\frac{\sin\left(2\theta_{\psi}\right)}{\sin\left(2\pi\Delta_{\psi}\right)},
\label{Q_psi}
\\
\mathcal{Q}_{\chi} &
=\frac{1}{2}-\frac{\theta_{\chi}}{\pi}-\left(\frac{1}{2}-\Delta_{\chi}\right)\frac{\sin\left(2\theta_{\chi}\right)}{\sin\left(2\pi\Delta_{\chi}\right)}.
\label{Q_chi}
\end{flalign}

The five equations above, $i.e.$ Eq.~\ref{self-consistent eq1} to
Eq.~\ref{Q_chi} involve five unknown real numbers that we need to
solve for: $\Delta_\psi$, $\Delta_\chi$, $\theta_\psi$,
$\theta_\chi$, and $C_{\psi}^2C_{\chi}^{2}$. These equations imply
that the conformal dimension $\Delta_{\psi/\chi}$ can be tuned by
the particle filling factors $\mathcal{Q}_{\psi}$ and
$\mathcal{Q}_{\chi}$, as we will demonstrate explicitly later.


The imaginary time correlation function can be obtained by Fourier
transforming Eq.~\ref{ansatz1} and Eq.~\ref{ansatz2}:
 \beqn
 G_{\psi/\chi}\left(\tau\right) =
 \frac{\mathcal{B}_{\psi/\chi}}{\left|\tau\right|^{2\Delta_{\psi/\chi}}},
 \ \ \
 \tau>0 \cr\cr
 G_{\psi/\chi}\left(\tau\right) =
 -\frac{\mathcal{B}'_{\psi/\chi}}{\left|\tau\right|^{2\Delta_{\psi/\chi}}}, \ \ \
 \tau<0,
 \eeqn
Following the convention of the literatures on the complex SYK
model (for example Ref.~\onlinecite{Sachdev2015}), we can
introduce the spectral asymmetry $\mathcal{E}_{\psi/\chi}$
 \begin{equation}
 e^{2\pi\mathcal{E}_{\psi/\chi}}=\frac{\sin\left(\pi\Delta_{\psi/\chi}+\theta_{\psi/\chi}\right)}{\sin\left(\pi\Delta_{\psi/\chi}-\theta_{\psi/\chi}\right)},
 \end{equation}
and the coefficient $\mathcal{B}_{\psi/\chi}$,
$\mathcal{B}_{\psi/\chi}'$ is related to
 $C_{\psi/\chi}$ as
 \beqn
 \mathcal{B}_{\psi/\chi} &=& -\frac{C_{\psi/\chi}\Gamma\left(2\Delta_{\psi/\chi}\right)\sin\left(\pi\Delta_{\psi/\chi}+\theta_{\psi/\chi}\right)}{\pi},
 \cr\cr
 \mathcal{B}'_{\psi/\chi} &=& -\frac{C_{\psi/\chi}\Gamma\left(2\Delta_{\psi/\chi}\right)\sin\left(\pi\Delta_{\psi/\chi} -
 \theta_{\psi/\chi}\right)}{\pi}
 \cr\cr
 &=& \mathcal{B}_{\psi/\chi}e^{ - 2\pi\mathcal{E}_{\psi/\chi}}.
 \eeqn Although we cannot determine $C_\psi$ and $C_\chi$
 separately from the S-D equations, dimensional analysis determines that $\mathcal{B}_{\psi/\chi} \sim C_{\psi/\chi} \sim
 g^{-2\Delta_{\psi/\chi}}$, thus $C_\psi^2 C_\chi^2 \sim 1/g^2 $.

The finite temperature solution can be obtained by performing the
conformal mapping $\tau\rightarrow\frac{1}{\pi T}\tan\left(\pi
T\tau\right)$, where $\tau$ becomes a periodic imaginary time
coordinate with periodicity $1/T$.  Using the rules of
reparametrization transformation, we obtain
 \begin{equation}
 G\left(\tau\right)=\begin{cases}
 \mathcal{B}e^{-2\pi\mathcal{E}T\tau}\left|\frac{\pi T}{\sin\left(\pi T\tau\right)}\right|^{2\Delta} & 0<\tau<\frac{1}{T}\\
 -\mathcal{B}' e^{-2\pi\mathcal{E}T\tau}\left|\frac{\pi T}{\sin\left(\pi T\tau\right)}\right|^{2\Delta} & 0<-\tau<\frac{1}{T}
 \end{cases},
 \end{equation}

Now we are ready to solve the equations from
Eq.~\ref{self-consistent eq1} to Eq.~\ref{Q_chi}. In general an
analytic solution would be very tedious. But for the simplified
case where $M_1 = M_2$, there are only two parameters in this
theory: $q_\psi = \mathcal{Q}_{\psi} - 1/2$ and $q_\chi =
\mathcal{Q}_{\chi} -1/2$, and all the relevant quantities can be
expanded as a polynomial of $q_\psi$, $q_\chi$. We also define $d
= \Delta_\psi - 1/4 = 1/4 - \Delta_\chi$. Then Eq.~\ref{Dcons}
implies that $d_\psi = - d_\chi = d$. We will obtain analytic
solutions for small $q_\psi$ and $q_\chi$.

In fact, in Eq.~\ref{Q_chi} and Eq.~\ref{Q_psi}, we do not need to
compute the exact prefactor before $\sin(2\theta_\psi)$ and
$\sin(2\theta_\chi)$. Without loss of generality, we can assume
the prefactor $f(\Delta, \theta)$ is a function of $\Delta$ and
$\theta$, and some general constraints of the its form would be
sufficient for the lowest nontrivial order of solutions as a
polynomial of $q_{\psi/\chi}$. For example, $f(\Delta, \theta)$
must be consistent with the results in
Ref.~\onlinecite{sachdev2001}. When $q_\psi = q_\chi$, there is a
$Z_2$ symmetry that exchanges $\psi$ and $\chi$, hence in this
case $\Delta_\psi = \Delta_\chi = 1/4$, or $d = d_\psi = - d_\chi
= 0$. And to be consistent with the result in
Ref.~\onlinecite{sachdev2001}, the $f(\Delta, \theta)$ function
must satisfy \beqn f(1/4, \theta) = 1/4, \eeqn and this statement
is independent of $\theta$. This is consistent with the result of
Ref.~\onlinecite{Gu2017b} where it was found that $f(\Delta,
\theta)$ does not depend on $\theta$ at all.

Under the particle-hole transformation, the Green's function
$G(\tau)$ at filling factor $q_\psi$, $q_\chi$ will become  $- G(
- \tau)$ at filling factor $- q_\psi$, $- q_\chi$. This implies
that $d$ must be an even function of $q_\psi$ and $q_\chi$, while
$\theta_\psi$, $\theta_\chi$ must be odd functions of $q_\psi$,
$q_\chi$. If we assume $q_\psi \sim q_\chi \sim q \ll 1$ , to the
lowest order expansion of $q_\psi$ and $q_\chi$, $d \sim (q_\psi^2
- q_\chi^2) $, which follows from the aforementioned fact that $d
= 0$ when $q_\psi = q_\chi$. Thus to the lowest nontrivial order
of expansion of $q$, we can just take $f(\Delta, \theta) = 1/4 +
O(q_\psi^2 - q_\chi^2) + O(q^3)$.

All the five equations from Eq.~\ref{self-consistent eq1} to
Eq.~\ref{Q_chi} can be expanded as a polynomial of $q_\psi$ and
$q_\chi$. And at the lowest nontrivial order, we obtain the
following analytic solutions: \beqn \theta_\psi &=& - \frac{2\pi
q_\psi}{\pi + 2} + O(q^3), \cr\cr \theta_\chi &=& - \frac{2\pi
q_\chi}{\pi + 2} + O(q^3), \cr\cr \Delta_{\psi} &=& \frac{1}{4} +
d = \frac{1}{4} + \frac{2\pi^2 (q_\psi^2 - q_\chi^2)}{(\pi+2)^2
(\pi - 2)} + O(q^4), \cr\cr \Delta_{\chi} &=&\frac{1}{4} - d =
\frac{1}{4} - \frac{2\pi^2 (q_\psi^2 - q_\chi^2)}{(\pi+2)^2 (\pi -
2)} + O(q^4). \eeqn These solutions are consistent with all the
previous observations, and also consistent with numerical
solutions of the equations

\subsection{Model B}

Another elementary model that we will start with is also
constructed with two orbitals of fermions, each with three
indices. The Hamiltonian takes the following form: \beqn H_{0}^B
&& = \sum_{a_1, a_2, b_1, b_2 = 1}^N \sum_{c, c' = 1}^{M_1}
\sum_{d, d' = 1}^{M_2} \frac{g}{N\sqrt{M}} \mathcal{J}^\psi_{c,
c'} \mathcal{J}^\chi_{d, d'} \cr\cr && \left( \psi^\dagger_{a_1,
b_1, c} \psi^\dagger_{a_2, b_2, c'} \chi_{a_1, b_2, d} \chi_{a_2,
b_1, d'} + h.c. \right), \label{H0B}\eeqn Here $\psi_c$ and
$\chi_d$ form fundamental representation of Sp($M_1$) and
Sp($M_2$) group. $\mathcal{J}^\psi_{c, c'} \psi_{c} \psi_{c'}$ and
$\mathcal{J}^\chi_{d, d'} \chi_{d} \chi_{d'}$ form singlets under
Sp($M_1$) and Sp($M_2$) respectively.

Although both model A and model B share a similar three-index
structure, there are some fundamental differences between them.
First of all, the particle density of $\psi$ and $\chi$ are no
longer separately conserved in model B. Only the total particle
density is conserved. Thus, we should introduce \beqn \mathcal{Q}=
\frac{M_1 \mathcal{Q}_{\psi} + M_2 \mathcal{Q}_{\chi}}{M_1 + M_2}
\in (0,1)  \label{total Q} \eeqn as a ``total'' filling factor,
Notice that $Q_\psi$ and $Q_\chi$ are defined as the expectation
values of $\psi$ and $\chi$ fermion number operator
(Eq.~\ref{ff}), while only $\mathcal{Q}$ is a conserved quantity
in this case.

Secondly and very importantly, the self energies are different
compared with those of model A, based on the melonic diagrams:
\begin{flalign}
 \Sigma_{\psi}^{B}\left(\tau\right) & =-4g^{2}\sqrt{\frac{M_{2}}{M_{1}}}
 G_{\chi}\left(\tau\right)^{2}G_{\psi}\left(-\tau\right),
 \\
 \Sigma_{\chi}^{B}\left(\tau\right) &
 =-4g^{2}\sqrt{\frac{M_{1}}{M_{2}}}G_{\psi}\left(\tau\right)^{2}G_{\chi}\left(-\tau\right),
\end{flalign}

Again, we want to solve the coupled S-D equations Eq.~\ref{S-D eq}
self-consistently in the conformal limit, and we still use the
power law ansatz Eq.~\ref{ansatz1} and Eq.~\ref{ansatz2}. We found
that the self energies $\tilde{\Sigma}_{\psi}^{B},
\tilde{\Sigma}_{\chi}^{B}$ can still be written as the form of
Eq.~\ref{proportional1}, Eq.~\ref{proportional2}. But now the
self-consistency of the S-D equation imposes another constraint on
$\theta_{\psi},\theta_{\chi}$ (for more details, please refer to
the supplementary material appendix A):
 \begin{flalign}
  \frac{\sin\left(\pi\Delta_\psi+\theta_{\psi}\right)}{\sin\left(\pi\Delta_\psi-\theta_{\psi}\right)}= \frac{\sin\left(\pi\Delta_\chi+\theta_{\chi}\right)}{\sin\left(\pi\Delta_\chi-\theta_{\chi}\right)},
 \label{constraint theta}
 \end{flalign}
 which implies that the two types of fermions have the same spectral asymmetry.
Under this constraint, the S-D equation Eq.~\ref{S-D eq} leads to
the same expressions as Eq.~\ref{self-consistent eq1} and
Eq.~\ref{self-consistent eq2}.

In addition, we have verified in the supplementary material that
the expectation values of the particle numbers for $\psi$ and
$\chi$ fermions share the same expressions Eq.~\ref{Q_psi} and
Eq.~\ref{Q_chi} as model A. The total filling factor $\mathcal{Q}$
imposes further constraints on $\Delta_{\psi/\chi}$, and
$\theta_{\psi/\chi}$
 \beqn
 \mathcal{Q} &=& \frac{1}{2} - \frac{M_1 \theta_{\psi}+ M_2 \theta_{\chi}}{\pi (M_1 + M_2)}
 \cr\cr
 &-& \frac{ M_1 \Delta_{\chi}\sin\left(2\theta_{\psi}\right) + M_2 \Delta_{\psi}\sin\left(2\theta_{\chi}\right)}{\sin\left(2\pi\Delta_{\psi/\chi}\right)(M_1 + M_2)},    \label{Q}
 \eeqn
where $\Delta_{\psi/\chi}$ can be either $\Delta_{\psi}$ or
$\Delta_{\chi}$ due to Eq.~\ref{Dcons}.

Still, we have five equations that involve five unknown real
quantities $\Delta_\psi, \Delta_\chi, \theta_\psi, \theta_\chi$,
and $C_{\psi}^2C_{\chi}^{2}$. Compared to model A, the conditions
that $\mathcal{Q}_{\psi}$ and $\mathcal{Q}_{\chi}$ are fixed
separately is replaced by fixing $\mathcal{Q}$, together with the
constraint Eq.~\ref{constraint theta}. Now the conformal dimension
$\Delta_{\psi/\chi}$ can be tuned by changing the total particle
filling factor $\mathcal{Q}$.

\section{ Lattice models for NFLs }

\begin{figure}
\includegraphics[scale=0.6]{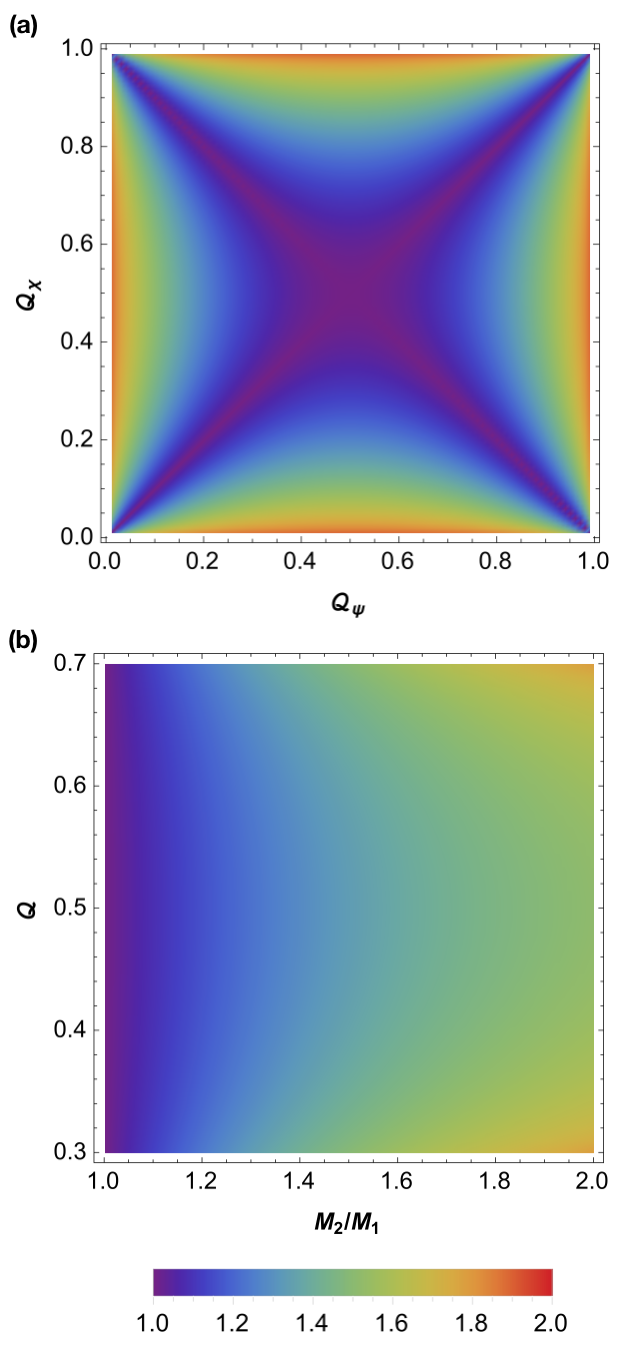}
\caption{The relation between the transport scaling power $\alpha$
(defined as resistivity $\varrho\sim T^{\alpha}$) and parameters in the lattice models for NFLs. $\left(a\right)$ $\alpha$ plotted against $\mathcal{Q}_{\psi}$ and $\mathcal{Q}_{\chi}$ with $M_{2}/M_{1}=1$ for the lattice model (1) with the on-cluster Hamiltonian $H_{0}^{A}\left(\boldsymbol{r}\right)$; $\left(b\right)$ $\alpha$ plotted against $\mathcal{Q}$ and $M_{2}/M_{1}$, for lattice model (1) with the on-cluster Hamiltonian $H_{0}^{B}\left(\boldsymbol{r}\right)$; and also the lattice model (2) Eq.~\ref{H0}.} \label{scaling}
\end{figure}

\subsection{Lattice model (1)}

Based on the elementary models constructed in the previous
section, we can construct lattice models with the desired
resistivity scaling $\varrho \sim T^\alpha$ with $\alpha \in
[1,2)$. Our first lattice model is constructed with coupled
clusters (following the previous
efforts~\cite{Song2017,Gu2017b,patel2017,berg2018} of constructing
the strange metal phase with the SYK$_4$-like clusters), and the
physics on each cluster $\boldsymbol{r}$ is described by
Eq.~\ref{H0A} or Eq.~\ref{H0B}, which is the leading energy scale
of the system. Different clusters are coupled together through
hoppings of both $\psi$ and $\chi$:
\begin{equation}
 H=\sum_{\boldsymbol{r}}H_{0}^{A/B}\left(\boldsymbol{r}\right)-\sum_{\left\langle
 \boldsymbol{r},\boldsymbol{r}^{\prime}\right\rangle
 }\left(t_{1}\psi_{\boldsymbol{r}}^{\dagger}\psi_{\boldsymbol{r}^{\prime}}+t_{2}\chi_{\boldsymbol{r}}^{\dagger}\chi_{\boldsymbol{r}^{\prime}}\right)+\ldots
 \label{lattice1}
\end{equation}
The indices of $\psi$ and $\chi$ are summed over in the equation
above. Although the $t$-terms are expected to drive the system
into a Fermi liquid state at low energy, our goal is to construct
a NFL phase {\it at a finite energy/temperature window}, which is
where most of the NFLs are observed experimentally. Thus let us
focus on the finite energy window where $H_{0}^{A/B}$ is dominant,
and the hopping term is perturbative.

The electric current operator of model Eq.~\ref{lattice1} can be
obtained by coupling the model to the external electromagnetic
field, and perform functional derivative of the external field:
 \begin{equation}
 J_{\boldsymbol{\delta}}=\sum_{\boldsymbol{r}}it_{1}\psi_{\boldsymbol{r}}^{\dagger}\psi_{\boldsymbol{r}+\boldsymbol{\delta}}+it_{2}\chi_{\boldsymbol{r}}^{\dagger}\chi_{\boldsymbol{r}+\boldsymbol{\delta}}+\textrm{H.c.}
 \end{equation}
In order to compute the electric conductivity, we define the
imaginary-time current-current correlation function as
$C\left(J,J;\tau\right)=\left\langle
\mathbb{T}_{\tau}J\left(\tau\right)J\left(0\right)\right\rangle $.
The leading order nonzero contribution takes the form
 \begin{equation}
 \frac{C\left(J,J;\tau\right)}{\mathcal{N}}=-2t_{1}^{2}G_{\psi}\left(\tau\right)G_{\psi}\left(-\tau\right)-2t_{2}^{2}G_{\chi}\left(\tau\right)G_{\chi}\left(-\tau\right),
 \end{equation}
where $\mathcal{N}$ is $\mathcal{N}=N^{2}M V$ with $V$ being the
size of the lattice.

Then we perform Fourier transformation of $C\left(J,J;\tau\right)$
to obtain correlation function in the Matsubara frequency space:
\begin{flalign}
\frac{C\left(J,J;i\omega_{n}\right)}{\mathcal{N}}=\frac{C_{\psi}\left(J,J;i\omega_{n}\right)}{\mathcal{N}}+\frac{C_{\chi}\left(J,J;i\omega_{n}\right)}{\mathcal{N}},
\end{flalign}
where $C_{\psi}$ is calculated as
\begin{flalign}
\frac{C_{\psi}\left(J,J;i\omega_{n}\right)}{\mathcal{N}}=2t_{1}^{2}\int_{0}^{\frac{1}{T}}d\tau e^{i\omega_{n}\tau}G_{\psi}\left(\tau\right)G_{\psi}\left(\frac{1}{T}-\tau\right), \label{C_JJ,psi}
\end{flalign} which is exact in the large$-N,M_1,M_2$ limit,
and $C_{\chi}$ has a similar expression.

When $0<\Delta_{\psi}<1/4$, the integral Eq.~\ref{C_JJ,psi} has a
finite expression, but it diverges when $1/4\leq\Delta_\psi<1/2$.
For $1/4\leq\Delta_\psi<1/2$, we regulate the integral by
introducing a small positive cutoff $\delta>0$:
\begin{equation}
 \int_{0}^{\frac{1}{T}}\rightarrow\int_{\delta}^{\frac{1}{T}-\delta}.
\end{equation}
There is a $\mathcal{O}\left(\log\delta\right)$ divergence when
$\Delta_\psi=1/4$, and a
$\mathcal{O}\left(1/\delta^{4\Delta-1}\right)$ divergence when
$1/4<\Delta_\psi<1/2$. The divergence is in the real part but not
the imaginary part of the correlation function, hence does not
contribute to the conductivity, thus the divergence can be removed
in order to calculate the conductivity.
The retarded/advanced correlation function
$C^{R/A}\left(J,J;\omega\right)$ can then be derived by taking $z
\rightarrow \omega\pm i0^{+}$. And eventually using the relation
$\sigma\left(\omega\right)=\frac{1}{i\omega}C^{R}\left(J,J;\omega\right)$,
we find the real part of the optical conductivity
 \begin{widetext}
 \begin{equation}
 \textrm{Re}[\sigma\left(\omega\right)]\sim\frac{t_{1}^{2}\mathcal{B}_{\psi}^{2}e^{-2\pi\mathcal{E}_{\psi}}}{T^{2-4\Delta_{\psi}}}\Upsilon\left(\Delta_{\psi},\frac{\omega}{T}\right)+\frac{t_{2}^{2}\mathcal{B}_{\chi}^{2}e^{-2\pi\mathcal{E}_{\chi}}}{T^{2-4\Delta_{\chi}}}\Upsilon\left(\Delta_{\chi},\frac{\omega}{T}\right), \label{conductivity}
 \end{equation}
where we have introduced the scaling function
 \begin{equation}
 \Upsilon\left(\Delta,\frac{\omega}{T}\right)=\frac{\left(2\pi\right)^{4\Delta-1}}{\Gamma\left(4\Delta\right)\cos\left(2\pi\Delta\right)}\frac{2\pi T}{\omega}\textrm{Im}\left[\frac{\Gamma\left(2\Delta+\frac{\omega}{i2\pi T}\right)}{\Gamma\left(1-2\Delta+\frac{\omega}{i2\pi T}\right)}\right]\qquad0<\Delta<1/2.
 \end{equation}
 \end{widetext}
One can check that when $\Delta=1/4$, the scaling function above
reproduces the scaling function for $\textrm{SYK}_{4}$-like
models~\cite{xu2018}
 \begin{equation}
 \Upsilon\left(\Delta=1/4,\omega/T\right)=\frac{\pi\tanh\left(\omega/2T\right)}{\omega/2T}.
 \end{equation}

The DC limit $\omega \rightarrow 0$ of the scaling function
$\Upsilon\left(\Delta, 0 \right)$ is a function of $\Delta$ which
takes finite positive values for $\Delta \in (0, 1/2)$. Since
$2\Delta_{\psi}+2\Delta_{\chi}=1$, the final result of the DC
conductivity takes the following form
 \begin{equation}
 \textrm{Re}[\sigma] \sim
 \frac{A}{T^{2-4\Delta}}+\frac{B}{T^{4\Delta}},
 \end{equation}
where $\Delta$ takes values in $0<\Delta<1/2$. The constants $A
\sim t_1^2 \mathcal{B}_\psi^2 \sim t_1^2/g^{4\Delta}$, and $B \sim
t_2^2 \mathcal{B}_\chi^2 \sim t_2^2 / g^{2 -4 \Delta}$. Hence when
$T < g$, the $A/T^{2-4\Delta}$ part of the DC conductivity will
dominate for $0<\Delta<1/4$, and $B/T^{4\Delta}$ dominates for
$1/4<\Delta<1/2$. Thus, in a finite temperature window for $T$
lower than the dominant energy scale $g$, and higher than the
infrared scale below which the hopping terms become
nonperturbative, we are able to realize non-fermi liquid behaviors
with resistivity $\varrho\sim T^{\alpha}$, and
$\alpha\in\left[1,2\right)$ depends on parameters in the theory,
especially the filling factors in the model.

The relation between $\alpha$ and the filling factors is plotted
in Fig.~\ref{scaling}. If we start with model A on every cluster,
$\alpha$ will depend on both $Q_\chi$ and $Q_\psi$ even when $M_1
= M_2$; if we start with model B, then $\alpha$ depends on the
total filling factor $Q$ when $M_1 \neq M_2$.

\subsection{Lattice model (2)}

In this section we we propose another different construction of
lattice model for NFL, by relating two of the three tensor indices
to the lattice site coordinates of a two dimensional square
lattice.

The dominant interaction in this model is
 \beqn H = \sum_{j} \sum_{r, r' = - (N-1)/2
 }^{(N-1)/2} \ \sum_{c,c' = 1}^{M_1} \sum_{d,d' =
 1}^{M_2} \frac{g}{N \sqrt{M}} \mathcal{J}^\psi_{c, c'}
 \mathcal{J}^\chi_{d, d'} \cr\cr \left( \psi^\dagger_{j_x, j_y,
 c} \psi^\dagger_{j_x + r, j_y + r', c'} \chi_{j_x, j_y +
 r', d} \chi_{j_x + r, j_y, d'} + h.c. \right). \label{H0}\eeqn
This Hamiltonian is motivated by and resembles $H_0^B$. $(j_x,
j_y)$ represents the $x$ and $y$ coordinates of the lattice site
$j$. Physically $\psi_c$ and $\chi_d$ can be thought of as two
types of fermions with $M_1 = 2J_1 + 1$ and $M_2 = 2J_2 + 1$ total
angular momentum components, and the Hamiltonian represents the
process of tunnelling between the pair singlets of $\chi$ and
$\psi$. The cluster model in the previous section is insensitive
to the spatial dimensions, while the construction of Eq.~\ref{H0}
most naturally applies to a two dimensional system.

In Eq.~\ref{H0}, we always take the thermodynamics limit first
(the sum of $j$ is taken on a square lattice with infinite size).
Then in the large-$N$ (in this model larger$-N$ means longer range
interaction) and large-$M_1,M_2$ limit, the fermion Green's
function is still dominated by the ``melonic diagrams'' and hence
the Schwinger-Dyson equations, and their solutions, remain the
same as model $H_0^B$. Notice that the single fermion Green's
function is completely local in space, which is guaranteed by the
fact that the Eq.~\ref{H0} conserves the center of mass.

In addition to the dominant interaction, we will also turn on a
single-particle hopping term as perturbations. Because
Eq.~\ref{H0} conserves the center of mass of the electrons, the
interaction Eq.~\ref{H0} alone cannot transport electric charge.
Thus the electric current operator only comes from the electron
hopping terms. In the soluble large-$\left(N, M_1, M_2\right)$
limit, we formally generalize the electric current operator to the
following form \beqn J_{x} &=& \frac{i t_1}{\sqrt{NM_1}}
 \left( \sum_{c} \psi_{j, c}^{\dagger}\psi_{j+
 \hat{x}, c} + \sqrt{\frac{N-1}{2}}
 \psi_{j, c}^{\dagger}\psi_{j + \hat{x} \pm \hat{y}, c} \right)
 \cr\cr &+& \frac{i t_2}{\sqrt{NM_2}} \left( \sum_{d}
 \chi_{j, d}^{\dagger}\chi_{j+ \hat{x}, d} + \sqrt{\frac{N-1}{2}}
 \chi_{j, d}^{\dagger}\chi_{j + \hat{x} \pm \hat{y}, d}
 \right) \cr\cr &+& H.c. \label{current}\eeqn
This electric current density can be derived by designing a
corresponding single-electron hopping term in the large-$\left(N,
M_1, M_2\right)$ limit (which involves both nearest- and
second-neighbor hopping) and coupling it to the external
electromagnetic field.

Using the large-$\left(N, M_1, M_2\right)$ solution of
Eq.~\ref{H0}, we can repeat all the calculations for conductivity
as we did for the previous model (1), and we arrive at the same
expression of conductivity Eq.~\ref{conductivity}. Thus, we again
have tunability of transport scalings within this construction.
The exponent $\alpha$ of $\varrho \sim T^\alpha$ is plotted
against the filling factor $Q$ and $M_2/M_1$ in
Fig.~\ref{scaling}$b$.

\section{Summary and discussion}

We constructed two examples of lattice models for non-fermi liquid
states whose DC resistivity scalings are tunable by adjusting the
charge density, which is a phenomenon observed in many physical
systems. Our lattice models are based on two versions of
``elementary'' models with {\it randomness free four fermion
interactions}, which are soluble in certain theoretical limit just
like the SYK model and the fermion tensor models. But unlike the
previous models, our elementary models have tunable fermion
scaling dimensions in their conformal solutions.

In this work we assumed that both orbitals (types) of the fermions
in the model carry electric charges. But at least for model A,
where the number of each type of fermions is conserved separately,
we can also assume that one of the two types of fermions are
charge neutral slave particles, which comes from
``fractionalizing'' the localized spins. This perspective is
similar to the the case in the original Sachdev-Ye
model~\cite{SachdevYe1993}, and also similar to a series of recent
studies~\cite{Gu2017b,patel2017,berg2018}. In this case, the slave
fermions will be coupled to a U(1) gauge field, whose effect in
the large$-N$ limit is expected to be suppressed, and the solution
of our model in the large$-N$ limit remains unchanged. In this
case the electric transport only comes from one of the two
orbitals of the fermions, and it is still tunable by changing the
charge density of the system.

In Ref.~\onlinecite{xu2017,xu2018}, it was shown that the SYK-type
of models are instable against extra marginally relevant
four-fermion interactions, and these perturbations can lead to
instability at low energy/temperature. In experiment, many of the
observed NFLs are preempted by ordered phases (for example
superconductivity) at low temperature. Also, it was shown in
Ref.~\onlinecite{sachdev2001} that the $1/N$ effect of the
original Sachdev-Ye model plays a role only at an exponentially
suppressed energy scale, and at finite temperature there is a wide
window where the conformal solution of the Sachdev-Ye model
applies. Similar effects were shown for the SYK model and also the
three-index tensor models by studying the subleading order of the
Feynmann diagrams~\cite{nextorder}. All these analysis can be
performed for our models as well, which we will defer to future
study.

The authors thank Yingfei Gu for very helpful discussions.
Chao-Ming Jian's research at KITP is supported by the Gordon and
Betty Moore Foundations EPiQS Initiative through Grant GBMF4304.
Cenke Xu is supported by the David and Lucile Packard Foundation.

\appendix

\begin{widetext}

\section{More details about Self energies}

\paragraph{Model A}

Using the S-D equations, the fermion self-energies
$\Sigma_{\psi/\chi}^{A}$ in imaginary time reads
\begin{flalign}
\tilde{\Sigma}_{\psi}^{A}\left(\tau\right) & =-2g^{2}C_{\psi}C_{\chi}^{2}\sqrt{\frac{M_{2}}{M_{1}}}\frac{\left(\cos\left(2\pi\Delta_{\psi}\right)+\cos\left(2\theta_{\chi}\right)\right)\Gamma\left(1-2\Delta_{\psi}\right)\sin\left(\pi\Delta_{\psi}+\textrm{sgn}\left(\tau\right)\theta_{\psi}\right)}{\pi^{2}\sin\left(2\pi\Delta_{\psi}\right)}\frac{\textrm{sgn}\left(\tau\right)}{\left|\tau\right|^{2-2\Delta_{\psi}}},\\
\tilde{\Sigma}_{\chi}^{A}\left(\tau\right) &
=-2g^{2}C_{\chi}C_{\psi}^{2}\sqrt{\frac{M_{1}}{M_{2}}}\frac{\left(\cos\left(2\pi\Delta_{\chi}\right)+\cos\left(2\theta_{\psi}\right)\right)\Gamma\left(1-2\Delta_{\chi}\right)\sin\left(\pi\Delta_{\chi}+\textrm{sgn}\left(\tau\right)\theta_{\chi}\right)}{\pi^{2}\sin\left(2\pi\Delta_{\chi}\right)}\frac{\textrm{sgn}\left(\tau\right)}{\left|\tau\right|^{2-2\Delta_{\chi}}}.
\end{flalign}
After Fourier transformation, the self-energy at complex frequency
$z,\textrm{Im}\left(z\right)>0$ reads
\begin{flalign}
\tilde{\Sigma}_{\psi}^{A}\left(z\right) & =-2g^{2}C_{\psi}C_{\chi}^{2}\sqrt{\frac{M_{2}}{M_{1}}}\frac{\cos\left(2\pi\Delta_{\psi}\right)+\cos\left(2\theta_{\chi}\right)}{\pi\left(1-2\Delta_{\psi}\right)\sin\left(2\pi\Delta_{\psi}\right)}e^{i\left(\pi\Delta_{\psi}+\theta_{\psi}\right)}z^{1-2\Delta_{\psi}},\\
\tilde{\Sigma}_{\chi}^{A}\left(z\right) &
=-2g^{2}C_{\chi}C_{\psi}^{2}\sqrt{\frac{M_{1}}{M_{2}}}\frac{\cos\left(2\pi\Delta_{\chi}\right)+\cos\left(2\theta_{\psi}\right)}{\pi\left(1-2\Delta_{\chi}\right)\sin\left(2\pi\Delta_{\chi}\right)}e^{i\left(\pi\Delta_{\chi}+\theta_{\chi}\right)}z^{1-2\Delta_{\chi}}.
\end{flalign}
We can see that the self-energy for model A automatically takes
the form $\Sigma^{A}\left(z\right) \propto
e^{i\left(\pi\Delta+\theta\right)}z^{1-2\Delta}$ with a real
factor.

\paragraph{Model B}

We then consider model B. Using the S-D equations, the
self-energies $\Sigma_{\psi/\chi}^{B}$ in imaginary time are
\begin{flalign}
\tilde{\Sigma}_{\psi}^{B}\left(\tau\right) & =-4g^{2}C_{\psi}C_{\chi}^{2}\sqrt{\frac{M_{2}}{M_{1}}}\frac{\cos^{2}\left(\pi\Delta_{\psi}-\textrm{sgn}\left(\tau\right)\theta_{\chi}\right)\sin\left(\pi\Delta_{\psi}-\textrm{sgn}\left(\tau\right)\theta_{\psi}\right)\Gamma\left(1-2\Delta_{\psi}\right)}{\pi^{2}\sin\left(2\pi\Delta_{\psi}\right)}\frac{\textrm{sgn}\left(\tau\right)}{\left|\tau\right|^{2-2\Delta_{\psi}}},\\
\tilde{\Sigma}_{\chi}^{B}\left(\tau\right) &
=-4g^{2}C_{\chi}C_{\psi}^{2}\sqrt{\frac{M_{1}}{M_{2}}}\frac{\cos^{2}\left(\pi\Delta_{\chi}-\textrm{sgn}\left(\tau\right)\theta_{\psi}\right)\sin\left(\pi\Delta_{\chi}-\textrm{sgn}\left(\tau\right)\theta_{\chi}\right)\Gamma\left(1-2\Delta_{\chi}\right)}{\pi^{2}\sin\left(2\pi\Delta_{\chi}\right)}\frac{\textrm{sgn}\left(\tau\right)}{\left|\tau\right|^{2-2\Delta_{\chi}}}.
\end{flalign}
Again, after Fourier transformation, the self-energy with
imaginary frequency reads:
\begin{flalign}
\tilde{\Sigma}_{\psi}^{B}\left(z\right) & =-g^{2}C_{\psi}C_{\chi}^{2}\sqrt{\frac{M_{2}}{M_{1}}}\frac{e^{-i2\left(\pi\Delta_{\psi}+\theta_{\chi}+\theta_{\psi}\right)}\left(\left(-1+e^{4i\theta_{\chi}}\right)e^{2i\left(\pi\Delta_{\psi}+\theta_{\psi}\right)}+2e^{2i\left(\pi\Delta_{\psi}+\theta_{\chi}\right)}+e^{4i\pi\Delta_{\psi}}+1\right)}{\pi\left(1-2\Delta_{\psi}\right)\sin\left(2\pi\Delta_{\psi}\right)}e^{i\left(\pi\Delta_{\psi}+\theta_{\psi}\right)}z^{1-2\Delta_{\psi}},\\
\tilde{\Sigma}_{\chi}^{B}\left(z\right) &
=-g^{2}C_{\chi}C_{\psi}^{2}\sqrt{\frac{M_{1}}{M_{2}}}\frac{e^{-i2\left(\pi\Delta_{\chi}+\theta_{\chi}+\theta_{\psi}\right)}\left(\left(-1+e^{4i\theta_{\psi}}\right)e^{2i\left(\pi\Delta_{\chi}+\theta_{\chi}\right)}+2e^{2i\left(\pi\Delta_{\chi}+\theta_{\psi}\right)}+e^{4i\pi\Delta_{\chi}}+1\right)}{\pi\left(1-2\Delta_{\chi}\right)\sin\left(2\pi\Delta_{\chi}\right)}e^{i\left(\pi\Delta_{\chi}+\theta_{\chi}\right)}z^{1-2\Delta_{\chi}}.
\end{flalign}
The self-consistency of the S-D equation demands the self-energy
take the form
$\Sigma^{B}\left(z\right)=-C^{-1}e^{i\left(\pi\Delta+\theta\right)}z^{1-2\Delta}$
with a real pre-factor $C$. Demanding the imaginary part of $C$
vanish leads to
\begin{flalign}
 \cos\left(\theta_{\chi}+\theta_{\psi}\right)\left(\sin^{2}\left(\pi\Delta_{\psi}\right)\sin\left(\theta_{\chi}\right)\cos\left(\theta_{\psi}\right)-\cos^{2}\left(\pi\Delta_{\psi}\right)\cos\left(\theta_{\chi}\right)\sin\left(\theta_{\psi}\right)\right)=0,\\
 \cos\left(\theta_{\chi}+\theta_{\psi}\right)\left(\sin^{2}\left(\pi\Delta_{\chi}\right)\sin\left(\theta_{\psi}\right)\cos\left(\theta_{\chi}\right)-\cos^{2}\left(\pi\Delta_{\chi}\right)\cos\left(\theta_{\psi}\right)\sin\left(\theta_{\chi}\right)\right)=0.
\end{flalign}
These equations can be simplified as
\begin{equation}
 \frac{\tan\left(\theta_{\psi}\right)}{\tan\left(\pi\Delta_{\psi}\right)}=\frac{\tan\left(\theta_{\chi}\right)}{\tan\left(\pi\Delta_{\chi}\right)},\label{eq:constraint_theta2p}
\end{equation}
where we have used $\Delta_{\psi}+\Delta_{\chi}=1/2$ to simplify
the equations. In fact, we can rewrite Eq. \ref{eq:constraint_theta2p} as
\begin{equation}
 \frac{\sin\left(\pi\Delta_\psi+\theta_{\psi}\right)}{\sin\left(\pi\Delta_\psi-\theta_{\psi}\right)}= \frac{\sin\left(\pi\Delta_\chi+\theta_{\chi}\right)}{\sin\left(\pi\Delta_\chi-\theta_{\chi}\right)},\label{eq:constraint_theta2}
\end{equation}
which implies that the two types of fermions have the same spectral asymmetry.

The S-D equation also requires
\begin{flalign}
 C_{\psi}^{-2}C_{\chi}^{-2} = 2g^{2}\sqrt{\frac{M_{2}}{M_{1}}}\frac{\cos\left(2\pi\Delta_{\psi}\right)\cos\left(2\left(\theta_{\chi}+\theta_{\psi}\right)\right)+\cos\left(2\theta_{\psi}\right)}{\pi\left(1-2\Delta_{\psi}\right)\sin\left(2\pi\Delta_{\psi}\right)},\\
 C_{\chi}^{-2}C_{\psi}^{-2} = 2g^{2}\sqrt{\frac{M_{1}}{M_{2}}}\frac{\cos\left(2\pi\Delta_{\chi}\right)\cos\left(2\left(\theta_{\chi}+\theta_{\psi}\right)\right)+\cos\left(2\theta_{\chi}\right)}{\pi\left(1-2\Delta_{\chi}\right)\sin\left(2\pi\Delta_{\chi}\right)}.
\end{flalign}
Imposing the constraints Eq.~\ref{constraint theta} or
Eq.~\ref{eq:constraint_theta2}, we recover exactly the same
self-consistent equations Eq.~\ref{self-consistent eq1} and
Eq.~\ref{self-consistent eq2} as the model A.

\section{Luttinger-Ward calculation}

Let us generalize the discussion by
Georges-Parcollet-Sachdev~\cite{sachdev2001} to our model, and the
goal is to establish the relation between the filling factors
(particle density) $\mathcal{Q}_{\psi},\mathcal{Q}_{\chi}$ of
model $A$, and $\mathcal{Q}$ of model $B$ to the most relevant
quantities such as $\Delta_{\psi/\chi}$ and $\theta_{\psi/\chi}$.

In the real-time formalism, at zero temperature, the filling
factor can be evaluated by computing the following
integral~\cite{sachdev2001}
 \begin{equation}
 i\mathbb{P}\int_{-\infty}^{+\infty}\frac{d\omega}{2\pi}e^{i\omega0^{+}}\left(\partial_{\omega}\log G\left(\omega\right)-G\left(\omega\right)\partial_{\omega}\tilde{\Sigma}\left(\omega\right)\right),
 \end{equation}
where
$G\left(\omega\right)=G^{R}\left(\omega\right)\varTheta\left(\omega\right)+G^{A}\left(\omega\right)\varTheta\left(-\omega\right)$
is the time-ordered Green function with
$\varTheta\left(\omega\right)$ being the Heaviside step function,
and $G^{R/A}\left(\omega\right)=G\left(\omega\pm i0^{+}\right)$ is
the real-time retarded/advanced Green's function obtained by
replacing $i\omega_{n}$ by $\omega\pm i0^{+}$ in the
imaginary-time Green's function. We use $\mathbb{P}$ to denote the
the principal value of the integral
$\mathbb{P}\int_{-\infty}^{+\infty}=\int_{-\infty}^{-\delta}+\int_{+\delta}^{+\infty}$
with a small positive cut off $\delta>0$~\cite{sachdev2001}.

Through the same line of arguments in Appendix A of
Ref.~\onlinecite{sachdev2001} (also see Appendix D of
Ref.~\onlinecite{Gu2017b}), the filling factors for both fermions
$\psi$ and $\chi$ are
 \begin{flalign}
 \mathcal{Q}_{\psi}&=\frac{1}{2}-\frac{\theta_{\psi}}{\pi}-i\mathbb{P}\int_{-\infty}^{+\infty}\frac{d\omega}{2\pi}e^{i\omega0^{+}}G_{\psi}\left(\omega\right)\partial_{\omega}\tilde{\Sigma}_{\psi}\left(\omega\right),\\\mathcal{Q}_{\chi}&=\frac{1}{2}-\frac{\theta_{\chi}}{\pi}-i\mathbb{P}\int_{-\infty}^{+\infty}\frac{d\omega}{2\pi}e^{i\omega0^{+}}G_{\chi}\left(\omega\right)\partial_{\omega}\tilde{\Sigma}_{\chi}\left(\omega\right).
 \end{flalign}
We are going to calculate the integral
\begin{equation}
\mathcal{I}_{\psi/\chi}^{A/B}=i\mathbb{P}\int_{-\infty}^{+\infty}\frac{d\omega}{2\pi}e^{i\omega0^{+}}G_{\psi/\chi}\left(\omega\right)\partial_{\omega}\tilde{\Sigma}_{\psi/\chi}^{A/B}\left(\omega\right) \label{integral I}
\end{equation}
for two fermions $\psi,\chi$ in both model $A$ and model $B$. To
do so, we will use the properties of the spectral functions
\begin{equation}
 \mathscr{A}_{\psi}\left(\omega\right)=\frac{C_{\psi}}{\pi}\frac{S_{\psi,\pm}}{\left|\omega\right|^{1-2\Delta_{\psi}}},\quad\mathscr{A}_{\chi}\left(\omega\right)=\frac{C_{\chi}}{\pi}\frac{S_{\chi,\pm}}{\left|\omega\right|^{1-2\Delta_{\chi}}},  \label{spectral function}
\end{equation}
where the notation $S_{\pm}$ stands for
$S_{\pm}=\sin\left(\pi\Delta\pm\theta\right)$, and $\pm$ depends
on the sign of $\omega$. Our convention here is
 \begin{equation}
 \mathscr{A}\left(\omega\right)=\mp\frac{1}{\pi}\mathrm{Im}G^{R/A}\left(\omega\right),\qquad G\left(z\right)=\int_{-\infty}^{+\infty}d\omega\frac{\mathscr{A}\left(\omega\right)}{z-\omega}.
 \end{equation}


\paragraph{Model A}
Using the melonic S-D equation, we obtain the Fourier
transformation of $\tilde{\Sigma}_{\psi}^{A}\left(\tau\right)$
 \begin{flalign}
\tilde{\Sigma}_{\psi}^{A}\left(\omega\right)=&-4g^{2}\sqrt{\frac{M_{2}}{M_{1}}}\int_{-\infty}^{+\infty}\frac{d\nu_{1}}{2\pi}\frac{d\nu_{2}}{2\pi}\frac{d\nu_{3}}{2\pi}G_{\psi}\left(\nu_{1}\right)G_{\chi}\left(\nu_{2}\right)G_{\chi}\left(\nu_{3}\right)2\pi\delta\left(\nu_{1}+\nu_{2}-\nu_{3}-\omega\right)\\=&-4g^{2}\sqrt{\frac{M_{2}}{M_{1}}}\int_{\left\{ \omega_{1}^{+},\omega_{2}^{+},\omega_{3}^{-}\right\} \cup\left\{ \omega_{1}^{-},\omega_{2}^{-},\omega_{3}^{+}\right\} }d\omega_{1}d\omega_{2}d\omega_{3}\frac{\mathscr{A}_{\psi}\left(\omega_{1}\right)\mathscr{A}_{\chi}\left(\omega_{2}\right)\mathscr{A}_{\chi}\left(\omega_{3}\right)}{\omega_{1}+\omega_{2}-\omega-\omega_{3}+i0^{+}\textrm{sgn}\left(\omega_{3}\right)},
 \end{flalign}
where the notation $\left\{
\omega_{1}^{+},\omega_{2}^{+},\omega_{3}^{-}\right\}$ means the
integration domain $\left\{
\omega_{1}>0,\omega_{2}>0,\omega_{3}<0\right\}$. Accordingly, the
integral Eq.~\ref{integral I} for $\psi$ reads
\begin{flalign}
\mathcal{I}_{\psi}^{A}&=i\mathbb{P}\int_{-\infty}^{+\infty}\frac{d\omega d\omega_{0}}{2\pi}\frac{\mathscr{A}_{\psi}\left(\omega_{0}\right)e^{i\omega0^{+}}}{\omega-\omega_{0}+i0^{+}\textrm{sgn}\left(\omega_{0}\right)}\partial_{\omega}\tilde{\Sigma}_{\psi}^{A}\left(\omega\right)\\&=\frac{4g^{2}}{2\pi i}\sqrt{\frac{M_{2}}{M_{1}}}\int_{\Gamma}d\omega_{0}d\omega_{1}d\omega_{2}d\omega_{3}\mathscr{A}_{\psi}\left(\omega_{0}\right)\mathscr{A}_{\psi}\left(\omega_{1}\right)\mathscr{A}_{\chi}\left(\omega_{2}\right)\mathscr{A}_{\chi}\left(\omega_{3}\right)\varPhi_{\delta}\left(\omega_{1}+\omega_{2}-\omega_{3}-i0^{+}\textrm{sgn}\omega_{1},\omega_{0}-i0^{+}\textrm{sgn}\omega_{0}\right).
 \end{flalign}
The integration domain of $\mathcal{I}_{\psi}^{A}$ is
$\Gamma=\Gamma_{1}\cup\Gamma_{2}\cup\Gamma_{3}\cup\Gamma_{4}$
where
\begin{flalign}
\Gamma_{1}=\left\{ \omega_{0}^{+},\omega_{1}^{+},\omega_{2}^{+},\omega_{3}^{-}\right\} ,\quad\Gamma_{2}=\left\{ \omega_{0}^{-},\omega_{1}^{+},\omega_{2}^{+},\omega_{3}^{-}\right\} ,\quad\Gamma_{3}=\left\{ \omega_{0}^{+},\omega_{1}^{-},\omega_{2}^{-},\omega_{3}^{+}\right\} ,\quad\Gamma_{4}=\left\{ \omega_{0}^{-},\omega_{1}^{-},\omega_{2}^{-},\omega_{3}^{+}\right\}.
 \end{flalign}
We have also used the function
 \begin{equation}
\varPhi_{\delta}\left(a+i\epsilon_{a},b+i\epsilon_{b}\right)=\mathbb{P}\int_{-\infty}^{+\infty}dz\frac{e^{i\omega0^{+}}}{\left(z-a-i\epsilon_{a}\right)^{2}\left(z-b-i\epsilon_{b}\right)}
\end{equation}
where $a,b\in\mathbb{R}$ and
$\epsilon_{a},\epsilon_{b}\rightarrow0$. The expression of
$\varPhi_{\delta}$ is explicitly calculated as Eq. A8 in
Ref.~\onlinecite{sachdev2001}. In the following, we will only use
its property
$\varPhi_{\delta}\left(-a-i\epsilon_{a},-b-i\epsilon_{b}\right)=-\varPhi_{\delta}\left(a+i\epsilon_{a},b+i\epsilon_{b}\right)$.
By changing of variables, we could write the integral as
\begin{flalign}
\mathcal{I}_{\psi}^{A}&=\frac{4g^{2}}{2\pi i}\sqrt{\frac{M_{2}}{M_{1}}}\int_{x_{i}>0}\prod_{i=0}^{3}dx_{i}\left(\begin{array}{c}
\left(\mathscr{A}_{\psi}\left(x_{1}\right)\mathscr{A}_{\chi}\left(x_{2}\right)\mathscr{A}_{\chi}\left(-x_{3}\right)\mathscr{A}_{\psi}\left(-x_{0}\right)-\mathscr{A}_{\psi}\left(-x_{1}\right)\mathscr{A}_{\chi}\left(-x_{2}\right)\mathscr{A}_{\chi}\left(x_{3}\right)\mathscr{A}_{\psi}\left(x_{0}\right)\right)\\
\times\varPhi_{\delta}\left(x_{1}+x_{2}+x_{3}-i\epsilon_{1},-x_{0}+i\epsilon_{0}\right)+\\
\left(\mathscr{A}_{\psi}\left(x_{1}\right)\mathscr{A}_{\chi}\left(x_{2}\right)\mathscr{A}_{\chi}\left(-x_{3}\right)\mathscr{A}_{\psi}\left(x_{0}\right)-\mathscr{A}_{\psi}\left(-x_{1}\right)\mathscr{A}_{\chi}\left(-x_{2}\right)\mathscr{A}_{\chi}\left(x_{3}\right)\mathscr{A}_{\psi}\left(-x_{0}\right)\right)\\
\times\varPhi_{\delta}\left(x_{1}+x_{2}+x_{3}-i\epsilon_{1},x_{0}-i\epsilon_{0}\right)
\end{array}\right).
 \end{flalign}

Using the expressions Eq.~\ref{spectral function}, we have
 \begin{flalign}
&\mathscr{A}_{\psi}\left(x_{1}\right)\mathscr{A}_{\chi}\left(x_{2}\right)\mathscr{A}_{\chi}\left(-x_{3}\right)\mathscr{A}_{\psi}\left(-x_{0}\right)-\mathscr{A}_{\psi}\left(-x_{1}\right)\mathscr{A}_{\chi}\left(-x_{2}\right)\mathscr{A}_{\chi}\left(x_{3}\right)\mathscr{A}_{\psi}\left(x_{0}\right)\\=&\frac{C_{\psi}^{2}C_{\chi}^{2}}{\pi^{4}}\frac{S_{\psi,+}S_{\chi,+}S_{\chi,-}S_{\psi,-}-S_{\psi,-}S_{\chi,-}S_{\chi,+}S_{\psi,+}}{\left|x_{0}\right|^{1-2\Delta_{\psi}}\left|x_{1}\right|^{1-2\Delta_{\psi}}\left|x_{2}\right|^{1-2\Delta_{\chi}}\left|x_{3}\right|^{1-2\Delta_{\chi}}}=0.
 \end{flalign}
 Thus, the first term vanishes, and we only need to calculate the second term
 \begin{flalign}
\mathcal{I}_{\psi}^{A}=\frac{4g^{2}}{2\pi i}\sqrt{\frac{M_{2}}{M_{1}}}\frac{C_{\psi}^{2}C_{\chi}^{2}}{\pi^{4}}\int_{u_{i}>0}\prod_{i=0}^{3}du_{i}\frac{S_{\psi,+}^{2}S_{\chi,+}S_{\chi,-}-S_{\psi,-}^{2}S_{\chi,-}S_{\chi,+}}{\left|u_{0}u_{1}\right|^{1-2\Delta_{\psi}}\left|u_{2}u_{3}\right|^{1-2\Delta_{\chi}}}\varPhi_{\delta=1}\left(u_{1}+u_{2}+u_{3}-i\epsilon_{1},u_{0}-i\epsilon_{0}\right),
 \end{flalign}
where we have introduced new variables $x_{i}=u_{i}\delta$ to take
the limit $\delta\rightarrow0^{+}$.

Before calculating the integral, we want to show
$\mathcal{I}_{\psi}^{A}$ does not depend on $M_1,M_2$. On one
hand, the straightforward calculation gives
 \begin{equation}
 S_{\psi,+}^{2}S_{\chi,+}S_{\chi,-}-S_{\psi,-}^{2}S_{\chi,-}S_{\chi,+}=\frac{1}{2}\sin\left(2\pi\Delta_{\psi}\right)\sin\left(2\theta_{\psi}\right)\left(\cos\left(2\theta_{\chi}\right)-\cos\left(2\pi\Delta_{\chi}\right)\right).
 \end{equation}
On the other hand, we read from the S-D equation
 \begin{equation}
 C_{\psi}^{2}C_{\chi}^{2}=\frac{1}{2g^{2}}\sqrt{\frac{M_{1}}{M_{2}}}\frac{\pi\left(1-2\Delta_{\psi}\right)\sin\left(2\pi\Delta_{\psi}\right)}{\cos\left(2\pi\Delta_{\psi}\right)+\cos\left(2\theta_{\chi}\right)}. \label{eq:C_psi^2 C_chi^2-1}
 \end{equation}
They together give us
 \begin{equation}
 \mathcal{I}_{\psi}^{A}=\frac{1}{i\pi^{4}}F^{A}\left(\Delta_{\psi}\right)\left(\frac{1}{2}-\Delta_{\psi}\right)\sin^{2}\left(2\pi\Delta_{\psi}\right)\sin\left(2\theta_{\psi}\right), \label{I^A_psi}
 \end{equation}
where
 \begin{equation}
 F^{A}\left(\Delta_{\psi}\right)=\int_{u_{i}>0}\prod_{i=0}^{3}du_{i}\frac{\varPhi_{\delta=1}\left(u_{1}+u_{2}+u_{3}-i\epsilon_{1},u_{0}-i\epsilon_{0}\right)}{\left|u_{0}u_{1}\right|^{1-2\Delta_{\psi}}\left|u_{2}u_{3}\right|^{2\Delta_{\psi}}}.
 \end{equation}

Then we define $x=u_{0},y=u_{1}+u_{2}+u_{3}$, and integrate over
$u_{2},u_{3}$. The result is
\begin{equation}
F\left(\Delta_{\psi}\right)=\frac{\pi}{\left(1-2\Delta_{\psi}\right)\sin\left(2\pi\Delta_{\psi}\right)}\int_{0}^{\infty}dxdy\left(\frac{y}{x}\right)^{1-2\Delta_{\psi}}\varPhi_{\delta=1}\left(y-i\epsilon_{1},x-i\epsilon_{0}\right). \label{F(Delta)}
\end{equation}
We proceed to calculate the integral in the following way
 \begin{flalign}
\int_{0}^{\infty}dxdy\left(\frac{y}{x}\right)^{1-2\Delta_{\psi}}\varPhi_{\delta=1}\left(y-i\epsilon_{1},x-i\epsilon_{0}\right)&=\int_{0}^{\infty}dxdy\left(\frac{y}{x}\right)^{1-2\Delta_{\psi}}\mathbb{P}_{\delta=1}\int_{-\infty}^{+\infty}dz\frac{e^{i\omega0^{+}}}{\left(z-y+i\epsilon_{1}\right)^{2}\left(z-x+i\epsilon_{0}\right)}\\&=\pi^{2}\frac{\left(1-2\Delta_{\psi}\right)}{\sin^{2}\left(2\pi\Delta_{\psi}\right)}\mathbb{P}_{\delta=1}\int_{-\infty}^{+\infty}dz\frac{e^{iz0^{+}}}{z}=i\pi^{3}\frac{\left(1-2\Delta_{\psi}\right)}{\sin^{2}\left(2\pi\Delta_{\psi}\right)}.
 \end{flalign}
 Thus, we have
\begin{equation}
 F\left(\Delta_{\psi}\right)=\frac{i\pi^{4}}{\sin^{3}\left(2\pi\Delta_{\psi}\right)}\quad\Longrightarrow\quad\mathcal{I}_{\psi}=\left(\frac{1}{2}-\Delta_{\psi}\right)\frac{\sin\left(2\theta_{\psi}\right)}{\sin\left(2\pi\Delta_{\psi}\right)}.
\end{equation}
In conclusion, we arrive at the result Eq.~\ref{Q_psi}, which is
consistent with the expression
$\mathcal{Q}\left(\theta,\Delta\right)$ in
Ref~\onlinecite{Gu2017b} for the complex $\textrm{SYK}_{q}$ model
with the conformal dimension $\Delta=1/q$.

Through similar calculations based on
\begin{flalign}
\mathcal{I}_{\chi}^{A}&=\frac{4g^{2}}{2\pi i}\sqrt{\frac{M_{1}}{M_{2}}}\frac{C_{\psi}^{2}C_{\chi}^{2}}{\pi^{4}}\int_{u_{i}>0}\prod_{i=0}^{3}du_{i}\frac{S_{\chi,+}^{2}S_{\psi,+}S_{\psi,-}-S_{\chi,-}^{2}S_{\psi,-}S_{\psi,+}}{\left|u_{0}u_{1}\right|^{1-2\Delta_{\psi}}\left|u_{2}u_{3}\right|^{1-2\Delta_{\chi}}}\varPhi_{\delta=1}\left(u_{1}+u_{2}+u_{3}-i\epsilon_{1},u_{0}-i\epsilon_{0}\right),
 \end{flalign}
we obtain the identical expression Eq.~\ref{Q_chi} for $\chi$
fermion. In model $A$, $\theta_{\psi},\theta_{\chi}$ are two
independent variables, and $\textrm{U}\left(1\right)$ charges for
$\psi,\chi$ are conserved separately.

\paragraph{Model B}

The expression of $\tilde{\Sigma}^{B}$ is a bit different from
$\tilde{\Sigma}^{A}$
\begin{flalign}
\tilde{\Sigma}_{\psi}^{B}\left(\omega\right)=&-4g^{2}\sqrt{\frac{M_{2}}{M_{1}}}\int_{-\infty}^{+\infty}\frac{d\nu_{1}}{2\pi}\frac{d\nu_{2}}{2\pi}\frac{d\nu_{3}}{2\pi}G_{\chi}\left(\nu_{1}\right)G_{\chi}\left(\nu_{2}\right)G_{\psi}\left(\nu_{3}\right)2\pi\delta\left(\nu_{1}+\nu_{2}-\nu_{3}-\omega\right)\\=&-4g^{2}\sqrt{\frac{M_{2}}{M_{1}}}\int_{\left\{ \omega_{1}^{+},\omega_{2}^{+},\omega_{3}^{-}\right\} \cup\left\{ \omega_{1}^{-},\omega_{2}^{-},\omega_{3}^{+}\right\} }d\omega_{1}d\omega_{2}d\omega_{3}\frac{\mathscr{A}_{\chi}\left(\omega_{1}\right)\mathscr{A}_{\chi}\left(\omega_{2}\right)\mathscr{A}_{\psi}\left(\omega_{3}\right)}{\omega_{1}+\omega_{2}-\omega-\omega_{3}+i0^{+}\textrm{sgn}\left(\omega_{3}\right)}.
 \end{flalign}
 Now  the integral Eq.~\ref{integral I} for $\psi$ reads
\begin{equation}
\mathcal{I}_{\psi}^{B}=\frac{4g^{2}}{2\pi i}\sqrt{\frac{M_{2}}{M_{1}}}\int_{\Gamma}d\omega_{0}d\omega_{1}d\omega_{2}d\omega_{3}\mathscr{A}_{\psi}\left(\omega_{0}\right)\mathscr{A}_{\chi}\left(\omega_{1}\right)\mathscr{A}_{\chi}\left(\omega_{2}\right)\mathscr{A}_{\psi}\left(\omega_{3}\right)\varPhi_{\delta}\left(\omega_{1}+\omega_{2}-\omega_{3}-i0^{+}\textrm{sgn}\omega_{1},\omega_{0}-i0^{+}\textrm{sgn}\omega_{0}\right)
\end{equation}
with the same integration domain as $\mathcal{I}_{\psi}^{A}$. By
changing of variables, we could write the integral as
\begin{flalign}
\mathcal{I}_{\psi}^{B}&=\frac{4g^{2}}{2\pi i}\sqrt{\frac{M_{2}}{M_{1}}}\int_{x_{i}>0}\prod_{i=0}^{3}dx_{i}\left(\begin{array}{c}
\left(\mathscr{A}_{\chi}\left(x_{1}\right)\mathscr{A}_{\chi}\left(x_{2}\right)\mathscr{A}_{\psi}\left(-x_{3}\right)\mathscr{A}_{\psi}\left(-x_{0}\right)-\mathscr{A}_{\chi}\left(-x_{1}\right)\mathscr{A}_{\chi}\left(-x_{2}\right)\mathscr{A}_{\psi}\left(x_{3}\right)\mathscr{A}_{\psi}\left(x_{0}\right)\right)\\
\times\varPhi_{\delta}\left(x_{1}+x_{2}+x_{3}-i\epsilon_{1},-x_{0}+i\epsilon_{0}\right)+\\
\left(\mathscr{A}_{\chi}\left(x_{1}\right)\mathscr{A}_{\chi}\left(x_{2}\right)\mathscr{A}_{\psi}\left(-x_{3}\right)\mathscr{A}_{\psi}\left(x_{0}\right)-\mathscr{A}_{\chi}\left(-x_{1}\right)\mathscr{A}_{\chi}\left(-x_{2}\right)\mathscr{A}_{\psi}\left(x_{3}\right)\mathscr{A}_{\psi}\left(-x_{0}\right)\right)\\
\times\varPhi_{\delta}\left(x_{1}+x_{2}+x_{3}-i\epsilon_{1},x_{0}-i\epsilon_{0}\right)
\end{array}\right).
 \end{flalign}
Using the expressions Eq.~\ref{spectral function}, we have
\begin{flalign}
 & \mathscr{A}_{\chi}\left(x_{1}\right)\mathscr{A}_{\chi}\left(x_{2}\right)\mathscr{A}_{\psi}\left(-x_{3}\right)\mathscr{A}_{\psi}\left(-x_{0}\right)-\mathscr{A}_{\chi}\left(-x_{1}\right)\mathscr{A}_{\chi}\left(-x_{2}\right)\mathscr{A}_{\psi}\left(x_{3}\right)\mathscr{A}_{\psi}\left(x_{0}\right)\\
= & \frac{C_{\psi}^{2}C_{\chi}^{2}}{\pi^{4}}\frac{S_{\chi,+}S_{\chi,+}S_{\psi,-}S_{\psi,-}-S_{\chi,-}S_{\chi,-}S_{\psi,+}S_{\psi,+}}{\left|x_{0}\right|^{1-2\Delta_{\psi}}\left|x_{1}\right|^{1-2\Delta_{\chi}}\left|x_{2}\right|^{1-2\Delta_{\chi}}\left|x_{3}\right|^{1-2\Delta_{\psi}}},
\end{flalign}
which seems nonzero at first glance. But it indeed vanishes due to
the constraint Eq.~\ref{constraint theta}, and we only need to
calculate the second term
\begin{equation}
\mathcal{I}_{\psi}^{B}=\frac{4g^{2}}{2\pi i}\sqrt{\frac{M_{2}}{M_{1}}}\frac{C_{\psi}^{2}C_{\chi}^{2}}{\pi^{4}}\int_{u_{i}>0}\prod_{i=0}^{3}du_{i}\frac{S_{\chi,+}^{2}S_{\psi,-}S_{\psi,+}-S_{\chi,-}^{2}S_{\psi,+}S_{\psi,-}}{\left|u_{0}u_{3}\right|^{1-2\Delta_{\psi}}\left|u_{1}u_{2}\right|^{1-2\Delta_{\chi}}}\varPhi_{\delta=1}\left(u_{1}+u_{2}+u_{3}-i\epsilon_{1},u_{0}-i\epsilon_{0}\right),
\end{equation}
where we have again used new variables $x_{i}=u_{i}\delta$. We
proceed to analyze the coefficient. The straightforward
calculation gives
\begin{equation}
S_{\chi,+}^{2}S_{\psi,+}S_{\psi,-}-S_{\chi,-}^{2}S_{\psi,-}S_{\psi,+}=\frac{1}{2}\sin\left(2\pi\Delta_{\chi}\right)\sin\left(2\theta_{\chi}\right)\left(\cos\left(2\theta_{\psi}\right)-\cos\left(2\pi\Delta_{\psi}\right)\right).
\end{equation}
By using the expression Eq.~\ref{eq:C_psi^2 C_chi^2-1} of
$C_{\psi}^{2}C_{\chi}^{2}$ and the constraint Eq.~\ref{constraint
theta}, we are able to obtain a similar form comparing to
Eq.~\ref{I^A_psi}
\begin{equation}
\mathcal{I}_{\psi}^{B}=\frac{1}{i\pi^{4}}F^{B}\left(\Delta_{\psi}\right)\left(\frac{1}{2}-\Delta_{\psi}\right)\sin^{2}\left(2\pi\Delta_{\psi}\right)\sin\left(2\theta_{\psi}\right),
\end{equation}
where
\begin{equation}
F^{B}\left(\Delta_{\psi}\right)=\int_{u_{i}>0}\prod_{i=0}^{3}du_{i}\frac{\varPhi_{\delta=1}\left(u_{1}+u_{2}+u_{3}-i\epsilon_{1},u_{0}-i\epsilon_{0}\right)}{\left|u_{0}u_{3}\right|^{1-2\Delta_{\psi}}\left|u_{1}u_{2}\right|^{2\Delta_{\psi}}}.
\end{equation}
The definition of $F^{B}\left(\Delta\right)$ here differs from
$F^{A}\left(\Delta\right)$ by exchanging $u_{1}\leftrightarrow
u_{3}$. Notice that $\epsilon_{1}=-\epsilon_{3}$ which makes the
definition looks nonequivalent. However, after defining
$x=u_{0},y=u_{1}+u_{2}+u_{3}$, and integrating over $u_{2},u_{3}$,
we still have the expression Eq.~\ref{F(Delta)}. Thus, we have
exactly the same result Eq.~\ref{Q_psi} for $\langle
\mathcal{Q}_{\psi}^{B} \rangle$.

Through similar calculations for $\chi$ fermion
\begin{flalign}
 \mathcal{I}_{\chi}^{B} & =\frac{4g^{2}}{2\pi i}\sqrt{\frac{M_{1}}{M_{2}}}\frac{C_{\psi}^{2}C_{\chi}^{2}}{\pi^{4}}\int_{u_{i}>0}\prod_{i=0}^{3}du_{i}\frac{S_{\psi,+}^{2}S_{\chi,-}S_{\chi,+}-S_{\psi,-}^{2}S_{\chi,+}S_{\chi,-}}{\left|u_{0}u_{3}\right|^{1-2\Delta_{\psi}}\left|u_{1}u_{2}\right|^{1-2\Delta_{\chi}}}\varPhi_{\delta=1}\left(u_{1}+u_{2}+u_{3}-i\epsilon_{1},u_{0}-i\epsilon_{0}\right),
\end{flalign}
we again obtain exactly the same expression Eq.~\ref{Q_chi} for
$\mathcal{Q}_{\chi}^{B}$. Despite the similarity in expressions,
only the total $\textrm{U}\left(1\right)$ charge filling factor
Eq.~\ref{total Q} is a conserved quantity in model B.

\end{widetext}

\bibliography{NFL}

\begin{thebibliography}{51}
\expandafter\ifx\csname natexlab\endcsname\relax\def\natexlab#1{#1}\fi
\expandafter\ifx\csname bibnamefont\endcsname\relax
  \def\bibnamefont#1{#1}\fi
\expandafter\ifx\csname bibfnamefont\endcsname\relax
  \def\bibfnamefont#1{#1}\fi
\expandafter\ifx\csname citenamefont\endcsname\relax
  \def\citenamefont#1{#1}\fi
\expandafter\ifx\csname url\endcsname\relax
  \def\url#1{\texttt{#1}}\fi
\expandafter\ifx\csname urlprefix\endcsname\relax\def\urlprefix{URL }\fi
\providecommand{\bibinfo}[2]{#2}
\providecommand{\eprint}[2][]{\url{#2}}

\bibitem[{\citenamefont{Hertz}(1976)}]{hertz}
\bibinfo{author}{\bibfnamefont{J.~A.} \bibnamefont{Hertz}},
  \bibinfo{journal}{Phys. Rev. B} \textbf{\bibinfo{volume}{14}},
  \bibinfo{pages}{1165} (\bibinfo{year}{1976}),
  \urlprefix\url{https://link.aps.org/doi/10.1103/PhysRevB.14.1165}.

\bibitem[{\citenamefont{Millis}(1993)}]{millis}
\bibinfo{author}{\bibfnamefont{A.~J.} \bibnamefont{Millis}},
  \bibinfo{journal}{Phys. Rev. B} \textbf{\bibinfo{volume}{48}},
  \bibinfo{pages}{7183} (\bibinfo{year}{1993}),
  \urlprefix\url{https://link.aps.org/doi/10.1103/PhysRevB.48.7183}.

\bibitem[{\citenamefont{L\"ohneysen
  et~al.}(2007{\natexlab{a}})\citenamefont{L\"ohneysen, Rosch, Vojta, and
  W\"olfle}}]{nfl1}
\bibinfo{author}{\bibfnamefont{H.~v.} \bibnamefont{L\"ohneysen}},
  \bibinfo{author}{\bibfnamefont{A.}~\bibnamefont{Rosch}},
  \bibinfo{author}{\bibfnamefont{M.}~\bibnamefont{Vojta}}, \bibnamefont{and}
  \bibinfo{author}{\bibfnamefont{P.}~\bibnamefont{W\"olfle}},
  \bibinfo{journal}{Rev. Mod. Phys.} \textbf{\bibinfo{volume}{79}},
  \bibinfo{pages}{1015} (\bibinfo{year}{2007}{\natexlab{a}}),
  \urlprefix\url{https://link.aps.org/doi/10.1103/RevModPhys.79.1015}.

\bibitem[{\citenamefont{Polchinski}(1994)}]{polchinskinfl}
\bibinfo{author}{\bibfnamefont{J.}~\bibnamefont{Polchinski}},
  \bibinfo{journal}{Nuclear Physics B} \textbf{\bibinfo{volume}{422}},
  \bibinfo{pages}{617 } (\bibinfo{year}{1994}), ISSN \bibinfo{issn}{0550-3213},
  \urlprefix\url{http://www.sciencedirect.com/science/article/pii/055032139490%
4499}.

\bibitem[{\citenamefont{Nayak and Wilczek}(1994{\natexlab{a}})}]{nayaknfl1}
\bibinfo{author}{\bibfnamefont{C.}~\bibnamefont{Nayak}} \bibnamefont{and}
  \bibinfo{author}{\bibfnamefont{F.}~\bibnamefont{Wilczek}},
  \bibinfo{journal}{Nuclear Physics B} \textbf{\bibinfo{volume}{417}},
  \bibinfo{pages}{359 } (\bibinfo{year}{1994}{\natexlab{a}}), ISSN
  \bibinfo{issn}{0550-3213},
  \urlprefix\url{http://www.sciencedirect.com/science/article/pii/055032139490%
4774}.

\bibitem[{\citenamefont{Nayak and Wilczek}(1994{\natexlab{b}})}]{nayaknfl2}
\bibinfo{author}{\bibfnamefont{C.}~\bibnamefont{Nayak}} \bibnamefont{and}
  \bibinfo{author}{\bibfnamefont{F.}~\bibnamefont{Wilczek}},
  \bibinfo{journal}{Nuclear Physics B} \textbf{\bibinfo{volume}{430}},
  \bibinfo{pages}{534 } (\bibinfo{year}{1994}{\natexlab{b}}), ISSN
  \bibinfo{issn}{0550-3213},
  \urlprefix\url{http://www.sciencedirect.com/science/article/pii/055032139490%
1589}.

\bibitem[{\citenamefont{Oganesyan et~al.}(2001)\citenamefont{Oganesyan,
  Kivelson, and Fradkin}}]{nematicnfl}
\bibinfo{author}{\bibfnamefont{V.}~\bibnamefont{Oganesyan}},
  \bibinfo{author}{\bibfnamefont{S.~A.} \bibnamefont{Kivelson}},
  \bibnamefont{and} \bibinfo{author}{\bibfnamefont{E.}~\bibnamefont{Fradkin}},
  \bibinfo{journal}{Phys. Rev. B} \textbf{\bibinfo{volume}{64}},
  \bibinfo{pages}{195109} (\bibinfo{year}{2001}),
  \urlprefix\url{https://link.aps.org/doi/10.1103/PhysRevB.64.195109}.

\bibitem[{\citenamefont{Lee}(2009)}]{nfl2}
\bibinfo{author}{\bibfnamefont{S.-S.} \bibnamefont{Lee}},
  \bibinfo{journal}{Phys. Rev. B} \textbf{\bibinfo{volume}{80}},
  \bibinfo{pages}{165102} (\bibinfo{year}{2009}),
  \urlprefix\url{https://link.aps.org/doi/10.1103/PhysRevB.80.165102}.

\bibitem[{\citenamefont{Mross et~al.}(2010)\citenamefont{Mross, McGreevy, Liu,
  and Senthil}}]{nfl3}
\bibinfo{author}{\bibfnamefont{D.~F.} \bibnamefont{Mross}},
  \bibinfo{author}{\bibfnamefont{J.}~\bibnamefont{McGreevy}},
  \bibinfo{author}{\bibfnamefont{H.}~\bibnamefont{Liu}}, \bibnamefont{and}
  \bibinfo{author}{\bibfnamefont{T.}~\bibnamefont{Senthil}},
  \bibinfo{journal}{Phys. Rev. B} \textbf{\bibinfo{volume}{82}},
  \bibinfo{pages}{045121} (\bibinfo{year}{2010}),
  \urlprefix\url{https://link.aps.org/doi/10.1103/PhysRevB.82.045121}.

\bibitem[{\citenamefont{Metlitski and Sachdev}(2010{\natexlab{a}})}]{nfl4}
\bibinfo{author}{\bibfnamefont{M.~A.} \bibnamefont{Metlitski}}
  \bibnamefont{and} \bibinfo{author}{\bibfnamefont{S.}~\bibnamefont{Sachdev}},
  \bibinfo{journal}{Phys. Rev. B} \textbf{\bibinfo{volume}{82}},
  \bibinfo{pages}{075127} (\bibinfo{year}{2010}{\natexlab{a}}),
  \urlprefix\url{https://link.aps.org/doi/10.1103/PhysRevB.82.075127}.

\bibitem[{\citenamefont{Metlitski and Sachdev}(2010{\natexlab{b}})}]{nfl5}
\bibinfo{author}{\bibfnamefont{M.~A.} \bibnamefont{Metlitski}}
  \bibnamefont{and} \bibinfo{author}{\bibfnamefont{S.}~\bibnamefont{Sachdev}},
  \bibinfo{journal}{Phys. Rev. B} \textbf{\bibinfo{volume}{82}},
  \bibinfo{pages}{075128} (\bibinfo{year}{2010}{\natexlab{b}}),
  \urlprefix\url{https://link.aps.org/doi/10.1103/PhysRevB.82.075128}.

\bibitem[{\citenamefont{Schlief et~al.}(2017)\citenamefont{Schlief, Lunts, and
  Lee}}]{nfl6}
\bibinfo{author}{\bibfnamefont{A.}~\bibnamefont{Schlief}},
  \bibinfo{author}{\bibfnamefont{P.}~\bibnamefont{Lunts}}, \bibnamefont{and}
  \bibinfo{author}{\bibfnamefont{S.-S.} \bibnamefont{Lee}},
  \bibinfo{journal}{Phys. Rev. X} \textbf{\bibinfo{volume}{7}},
  \bibinfo{pages}{021010} (\bibinfo{year}{2017}),
  \urlprefix\url{https://link.aps.org/doi/10.1103/PhysRevX.7.021010}.

\bibitem[{\citenamefont{Schattner et~al.}(2016)\citenamefont{Schattner,
  Lederer, Kivelson, and Berg}}]{nflqmc}
\bibinfo{author}{\bibfnamefont{Y.}~\bibnamefont{Schattner}},
  \bibinfo{author}{\bibfnamefont{S.}~\bibnamefont{Lederer}},
  \bibinfo{author}{\bibfnamefont{S.~A.} \bibnamefont{Kivelson}},
  \bibnamefont{and} \bibinfo{author}{\bibfnamefont{E.}~\bibnamefont{Berg}},
  \bibinfo{journal}{Phys. Rev. X} \textbf{\bibinfo{volume}{6}},
  \bibinfo{pages}{031028} (\bibinfo{year}{2016}),
  \urlprefix\url{https://link.aps.org/doi/10.1103/PhysRevX.6.031028}.

\bibitem[{\citenamefont{Gurvitch and Fiory}(1987)}]{linear1}
\bibinfo{author}{\bibfnamefont{M.}~\bibnamefont{Gurvitch}} \bibnamefont{and}
  \bibinfo{author}{\bibfnamefont{A.~T.} \bibnamefont{Fiory}},
  \bibinfo{journal}{Phys. Rev. Lett.} \textbf{\bibinfo{volume}{59}},
  \bibinfo{pages}{1337} (\bibinfo{year}{1987}),
  \urlprefix\url{https://link.aps.org/doi/10.1103/PhysRevLett.59.1337}.

\bibitem[{\citenamefont{Tozer et~al.}(1987)\citenamefont{Tozer, Kleinsasser,
  Penney, Kaiser, and Holtzberg}}]{linear2}
\bibinfo{author}{\bibfnamefont{S.~W.} \bibnamefont{Tozer}},
  \bibinfo{author}{\bibfnamefont{A.~W.} \bibnamefont{Kleinsasser}},
  \bibinfo{author}{\bibfnamefont{T.}~\bibnamefont{Penney}},
  \bibinfo{author}{\bibfnamefont{D.}~\bibnamefont{Kaiser}}, \bibnamefont{and}
  \bibinfo{author}{\bibfnamefont{F.}~\bibnamefont{Holtzberg}},
  \bibinfo{journal}{Phys. Rev. Lett.} \textbf{\bibinfo{volume}{59}},
  \bibinfo{pages}{1768} (\bibinfo{year}{1987}),
  \urlprefix\url{https://link.aps.org/doi/10.1103/PhysRevLett.59.1768}.

\bibitem[{\citenamefont{Martin et~al.}(1988)\citenamefont{Martin, Fiory,
  Fleming, Schneemeyer, and Waszczak}}]{linear3}
\bibinfo{author}{\bibfnamefont{S.}~\bibnamefont{Martin}},
  \bibinfo{author}{\bibfnamefont{A.~T.} \bibnamefont{Fiory}},
  \bibinfo{author}{\bibfnamefont{R.~M.} \bibnamefont{Fleming}},
  \bibinfo{author}{\bibfnamefont{L.~F.} \bibnamefont{Schneemeyer}},
  \bibnamefont{and} \bibinfo{author}{\bibfnamefont{J.~V.}
  \bibnamefont{Waszczak}}, \bibinfo{journal}{Phys. Rev. Lett.}
  \textbf{\bibinfo{volume}{60}}, \bibinfo{pages}{2194} (\bibinfo{year}{1988}),
  \urlprefix\url{https://link.aps.org/doi/10.1103/PhysRevLett.60.2194}.

\bibitem[{\citenamefont{Martin et~al.}(1990)\citenamefont{Martin, Fiory,
  Fleming, Schneemeyer, and Waszczak}}]{linear4}
\bibinfo{author}{\bibfnamefont{S.}~\bibnamefont{Martin}},
  \bibinfo{author}{\bibfnamefont{A.~T.} \bibnamefont{Fiory}},
  \bibinfo{author}{\bibfnamefont{R.~M.} \bibnamefont{Fleming}},
  \bibinfo{author}{\bibfnamefont{L.~F.} \bibnamefont{Schneemeyer}},
  \bibnamefont{and} \bibinfo{author}{\bibfnamefont{J.~V.}
  \bibnamefont{Waszczak}}, \bibinfo{journal}{Phys. Rev. B}
  \textbf{\bibinfo{volume}{41}}, \bibinfo{pages}{846} (\bibinfo{year}{1990}),
  \urlprefix\url{https://link.aps.org/doi/10.1103/PhysRevB.41.846}.

\bibitem[{\citenamefont{Varma et~al.}(1989)\citenamefont{Varma, Littlewood,
  Schmitt-Rink, Abrahams, and Ruckenstein}}]{Varma1989}
\bibinfo{author}{\bibfnamefont{C.~M.} \bibnamefont{Varma}},
  \bibinfo{author}{\bibfnamefont{P.~B.} \bibnamefont{Littlewood}},
  \bibinfo{author}{\bibfnamefont{S.}~\bibnamefont{Schmitt-Rink}},
  \bibinfo{author}{\bibfnamefont{E.}~\bibnamefont{Abrahams}}, \bibnamefont{and}
  \bibinfo{author}{\bibfnamefont{A.~E.} \bibnamefont{Ruckenstein}},
  \bibinfo{journal}{Phys. Rev. Lett.} \textbf{\bibinfo{volume}{63}},
  \bibinfo{pages}{1996} (\bibinfo{year}{1989}),
  \urlprefix\url{https://link.aps.org/doi/10.1103/PhysRevLett.63.1996}.

\bibitem[{\citenamefont{Cao et~al.}(2019)\citenamefont{Cao, Chowdhury,
  Rodan-Legrain, Rubies-Bigordà, Watanabe, Taniguchi, Senthil, and
  Jarillo-Herrero}}]{pablostrange}
\bibinfo{author}{\bibfnamefont{Y.}~\bibnamefont{Cao}},
  \bibinfo{author}{\bibfnamefont{D.}~\bibnamefont{Chowdhury}},
  \bibinfo{author}{\bibfnamefont{D.}~\bibnamefont{Rodan-Legrain}},
  \bibinfo{author}{\bibfnamefont{O.}~\bibnamefont{Rubies-Bigordà}},
  \bibinfo{author}{\bibfnamefont{K.}~\bibnamefont{Watanabe}},
  \bibinfo{author}{\bibfnamefont{T.}~\bibnamefont{Taniguchi}},
  \bibinfo{author}{\bibfnamefont{T.}~\bibnamefont{Senthil}}, \bibnamefont{and}
  \bibinfo{author}{\bibfnamefont{P.}~\bibnamefont{Jarillo-Herrero}},
  \bibinfo{journal}{arXiv:1901.03710}  (\bibinfo{year}{2019}).

\bibitem[{\citenamefont{{Sachdev} and {Ye}}(1993)}]{SachdevYe1993}
\bibinfo{author}{\bibfnamefont{S.}~\bibnamefont{{Sachdev}}} \bibnamefont{and}
  \bibinfo{author}{\bibfnamefont{J.}~\bibnamefont{{Ye}}},
  \bibinfo{journal}{Physical Review Letters} \textbf{\bibinfo{volume}{70}},
  \bibinfo{pages}{3339} (\bibinfo{year}{1993}), \eprint{cond-mat/9212030}.

\bibitem[{\citenamefont{{Kitaev}}(2015)}]{Kitaev2015}
\bibinfo{author}{\bibfnamefont{A.}~\bibnamefont{{Kitaev}}},
  \emph{\bibinfo{title}{{A simple model of quantum holography}}},
  \bibinfo{howpublished}{\url{http://online.kitp.ucsb.edu/online/entangled15/k%
itaev/,http: //online.kitp.ucsb.edu/online/entangled15/kitaev2/.}}
  (\bibinfo{year}{2015}), \bibinfo{note}{{T}alks at KITP, April 7, 2015 and May
  27, 2015.}

\bibitem[{\citenamefont{{Sachdev}}(2015)}]{Sachdev2015}
\bibinfo{author}{\bibfnamefont{S.}~\bibnamefont{{Sachdev}}},
  \bibinfo{journal}{Physical Review X} \textbf{\bibinfo{volume}{5}},
  \bibinfo{eid}{041025} (\bibinfo{year}{2015}), \eprint{1506.05111}.

\bibitem[{\citenamefont{{Polchinski} and {Rosenhaus}}(2016)}]{Polchinski2016}
\bibinfo{author}{\bibfnamefont{J.}~\bibnamefont{{Polchinski}}}
  \bibnamefont{and}
  \bibinfo{author}{\bibfnamefont{V.}~\bibnamefont{{Rosenhaus}}},
  \bibinfo{journal}{Journal of High Energy Physics}
  \textbf{\bibinfo{volume}{4}}, \bibinfo{eid}{1} (\bibinfo{year}{2016}),
  \eprint{1601.06768}.

\bibitem[{\citenamefont{{Maldacena} and
  {Stanford}}(2016)}]{MaldacenaStanford2016}
\bibinfo{author}{\bibfnamefont{J.}~\bibnamefont{{Maldacena}}} \bibnamefont{and}
  \bibinfo{author}{\bibfnamefont{D.}~\bibnamefont{{Stanford}}},
  \bibinfo{journal}{\prd} \textbf{\bibinfo{volume}{94}}, \bibinfo{eid}{106002}
  (\bibinfo{year}{2016}), \eprint{1604.07818}.

\bibitem[{\citenamefont{{Witten}}(2016)}]{Witten2016}
\bibinfo{author}{\bibfnamefont{E.}~\bibnamefont{{Witten}}},
  \bibinfo{journal}{ArXiv e-prints}  (\bibinfo{year}{2016}),
  \eprint{1610.09758}.

\bibitem[{\citenamefont{Klebanov and Tarnopolsky}(2017)}]{Klebanov2016}
\bibinfo{author}{\bibfnamefont{I.~R.} \bibnamefont{Klebanov}} \bibnamefont{and}
  \bibinfo{author}{\bibfnamefont{G.}~\bibnamefont{Tarnopolsky}},
  \bibinfo{journal}{Phys. Rev. D} \textbf{\bibinfo{volume}{95}},
  \bibinfo{pages}{046004} (\bibinfo{year}{2017}),
  \urlprefix\url{https://link.aps.org/doi/10.1103/PhysRevD.95.046004}.

\bibitem[{\citenamefont{Gross and Rosenhaus}(2017)}]{Gross2017}
\bibinfo{author}{\bibfnamefont{D.~J.} \bibnamefont{Gross}} \bibnamefont{and}
  \bibinfo{author}{\bibfnamefont{V.}~\bibnamefont{Rosenhaus}},
  \bibinfo{journal}{Journal of High Energy Physics}
  \textbf{\bibinfo{volume}{2017}}, \bibinfo{pages}{93} (\bibinfo{year}{2017}),
  ISSN \bibinfo{issn}{1029-8479},
  \urlprefix\url{https://doi.org/10.1007/JHEP02(2017)093}.

\bibitem[{\citenamefont{Bi et~al.}(2017)\citenamefont{Bi, Jian, You, Pawlak,
  and Xu}}]{xu2017}
\bibinfo{author}{\bibfnamefont{Z.}~\bibnamefont{Bi}},
  \bibinfo{author}{\bibfnamefont{C.-M.} \bibnamefont{Jian}},
  \bibinfo{author}{\bibfnamefont{Y.-Z.} \bibnamefont{You}},
  \bibinfo{author}{\bibfnamefont{K.~A.} \bibnamefont{Pawlak}},
  \bibnamefont{and} \bibinfo{author}{\bibfnamefont{C.}~\bibnamefont{Xu}},
  \bibinfo{journal}{Physical Review B} \textbf{\bibinfo{volume}{95}},
  \bibinfo{pages}{205105} (\bibinfo{year}{2017}).

\bibitem[{\citenamefont{Luo et~al.}(2017)\citenamefont{Luo, You, Li, Jian, Lu,
  Xu, Zeng, and Laflamme}}]{pairex}
\bibinfo{author}{\bibfnamefont{Z.}~\bibnamefont{Luo}},
  \bibinfo{author}{\bibfnamefont{Y.-Z.} \bibnamefont{You}},
  \bibinfo{author}{\bibfnamefont{J.}~\bibnamefont{Li}},
  \bibinfo{author}{\bibfnamefont{C.-M.} \bibnamefont{Jian}},
  \bibinfo{author}{\bibfnamefont{D.}~\bibnamefont{Lu}},
  \bibinfo{author}{\bibfnamefont{C.}~\bibnamefont{Xu}},
  \bibinfo{author}{\bibfnamefont{B.}~\bibnamefont{Zeng}}, \bibnamefont{and}
  \bibinfo{author}{\bibfnamefont{R.}~\bibnamefont{Laflamme}},
  \bibinfo{journal}{arXiv:1712.06458}  (\bibinfo{year}{2017}).

\bibitem[{\citenamefont{Metlitski et~al.}(2015)\citenamefont{Metlitski, Mross,
  Sachdev, and Senthil}}]{nflpairing1}
\bibinfo{author}{\bibfnamefont{M.~A.} \bibnamefont{Metlitski}},
  \bibinfo{author}{\bibfnamefont{D.~F.} \bibnamefont{Mross}},
  \bibinfo{author}{\bibfnamefont{S.}~\bibnamefont{Sachdev}}, \bibnamefont{and}
  \bibinfo{author}{\bibfnamefont{T.}~\bibnamefont{Senthil}},
  \bibinfo{journal}{Phys. Rev. B} \textbf{\bibinfo{volume}{91}},
  \bibinfo{pages}{115111} (\bibinfo{year}{2015}),
  \urlprefix\url{https://link.aps.org/doi/10.1103/PhysRevB.91.115111}.

\bibitem[{\citenamefont{Wang and Chubukov}(2015)}]{nflpairing2}
\bibinfo{author}{\bibfnamefont{Y.}~\bibnamefont{Wang}} \bibnamefont{and}
  \bibinfo{author}{\bibfnamefont{A.~V.} \bibnamefont{Chubukov}},
  \bibinfo{journal}{Phys. Rev. B} \textbf{\bibinfo{volume}{92}},
  \bibinfo{pages}{125108} (\bibinfo{year}{2015}),
  \urlprefix\url{https://link.aps.org/doi/10.1103/PhysRevB.92.125108}.

\bibitem[{\citenamefont{Mandal}(2016)}]{nflpairing3}
\bibinfo{author}{\bibfnamefont{I.}~\bibnamefont{Mandal}},
  \bibinfo{journal}{Phys. Rev. B} \textbf{\bibinfo{volume}{94}},
  \bibinfo{pages}{115138} (\bibinfo{year}{2016}),
  \urlprefix\url{https://link.aps.org/doi/10.1103/PhysRevB.94.115138}.

\bibitem[{\citenamefont{Lederer et~al.}(2015)\citenamefont{Lederer, Schattner,
  Berg, and Kivelson}}]{nflpairing4}
\bibinfo{author}{\bibfnamefont{S.}~\bibnamefont{Lederer}},
  \bibinfo{author}{\bibfnamefont{Y.}~\bibnamefont{Schattner}},
  \bibinfo{author}{\bibfnamefont{E.}~\bibnamefont{Berg}}, \bibnamefont{and}
  \bibinfo{author}{\bibfnamefont{S.~A.} \bibnamefont{Kivelson}},
  \bibinfo{journal}{Phys. Rev. Lett.} \textbf{\bibinfo{volume}{114}},
  \bibinfo{pages}{097001} (\bibinfo{year}{2015}),
  \urlprefix\url{https://link.aps.org/doi/10.1103/PhysRevLett.114.097001}.

\bibitem[{\citenamefont{Wang et~al.}(2016)\citenamefont{Wang, Abanov,
  Altshuler, Yuzbashyan, and Chubukov}}]{nflpairing5}
\bibinfo{author}{\bibfnamefont{Y.}~\bibnamefont{Wang}},
  \bibinfo{author}{\bibfnamefont{A.}~\bibnamefont{Abanov}},
  \bibinfo{author}{\bibfnamefont{B.~L.} \bibnamefont{Altshuler}},
  \bibinfo{author}{\bibfnamefont{E.~A.} \bibnamefont{Yuzbashyan}},
  \bibnamefont{and} \bibinfo{author}{\bibfnamefont{A.~V.}
  \bibnamefont{Chubukov}}, \bibinfo{journal}{Phys. Rev. Lett.}
  \textbf{\bibinfo{volume}{117}}, \bibinfo{pages}{157001}
  (\bibinfo{year}{2016}),
  \urlprefix\url{https://link.aps.org/doi/10.1103/PhysRevLett.117.157001}.

\bibitem[{\citenamefont{Lederer et~al.}(2017)\citenamefont{Lederer, Schattner,
  Berg, and Kivelson}}]{nflpairing6}
\bibinfo{author}{\bibfnamefont{S.}~\bibnamefont{Lederer}},
  \bibinfo{author}{\bibfnamefont{Y.}~\bibnamefont{Schattner}},
  \bibinfo{author}{\bibfnamefont{E.}~\bibnamefont{Berg}}, \bibnamefont{and}
  \bibinfo{author}{\bibfnamefont{S.~A.} \bibnamefont{Kivelson}},
  \bibinfo{journal}{Proceedings of the National Academy of Sciences}
  \textbf{\bibinfo{volume}{114}}, \bibinfo{pages}{4905} (\bibinfo{year}{2017}),
  \eprint{http://www.pnas.org/content/114/19/4905.full.pdf},
  \urlprefix\url{http://www.pnas.org/content/114/19/4905.abstract}.

\bibitem[{\citenamefont{Fradkin et~al.}(2015)\citenamefont{Fradkin, Kivelson,
  and Tranquada}}]{inter}
\bibinfo{author}{\bibfnamefont{E.}~\bibnamefont{Fradkin}},
  \bibinfo{author}{\bibfnamefont{S.~A.} \bibnamefont{Kivelson}},
  \bibnamefont{and} \bibinfo{author}{\bibfnamefont{J.~M.}
  \bibnamefont{Tranquada}}, \bibinfo{journal}{Rev. Mod. Phys.}
  \textbf{\bibinfo{volume}{87}}, \bibinfo{pages}{457} (\bibinfo{year}{2015}),
  \urlprefix\url{https://link.aps.org/doi/10.1103/RevModPhys.87.457}.

\bibitem[{\citenamefont{Mitrano et~al.}(2018)\citenamefont{Mitrano, Husain,
  Vig, Kogar, Rak, Rubeck, Schmalian, Uchoa, Schneeloch, Zhong
  et~al.}}]{abbamonte}
\bibinfo{author}{\bibfnamefont{M.}~\bibnamefont{Mitrano}},
  \bibinfo{author}{\bibfnamefont{A.~A.} \bibnamefont{Husain}},
  \bibinfo{author}{\bibfnamefont{S.}~\bibnamefont{Vig}},
  \bibinfo{author}{\bibfnamefont{A.}~\bibnamefont{Kogar}},
  \bibinfo{author}{\bibfnamefont{M.~S.} \bibnamefont{Rak}},
  \bibinfo{author}{\bibfnamefont{S.~I.} \bibnamefont{Rubeck}},
  \bibinfo{author}{\bibfnamefont{J.}~\bibnamefont{Schmalian}},
  \bibinfo{author}{\bibfnamefont{B.}~\bibnamefont{Uchoa}},
  \bibinfo{author}{\bibfnamefont{J.}~\bibnamefont{Schneeloch}},
  \bibinfo{author}{\bibfnamefont{R.}~\bibnamefont{Zhong}},
  \bibnamefont{et~al.}, \bibinfo{journal}{Proceedings of the National Academy
  of Sciences} \textbf{\bibinfo{volume}{115}}, \bibinfo{pages}{5392}
  (\bibinfo{year}{2018}), ISSN \bibinfo{issn}{0027-8424},
  \eprint{http://www.pnas.org/content/115/21/5392.full.pdf},
  \urlprefix\url{http://www.pnas.org/content/115/21/5392}.

\bibitem[{\citenamefont{Song et~al.}(2017)\citenamefont{Song, Jian, and
  Balents}}]{Song2017}
\bibinfo{author}{\bibfnamefont{X.-Y.} \bibnamefont{Song}},
  \bibinfo{author}{\bibfnamefont{C.-M.} \bibnamefont{Jian}}, \bibnamefont{and}
  \bibinfo{author}{\bibfnamefont{L.}~\bibnamefont{Balents}},
  \bibinfo{journal}{Phys. Rev. Lett.} \textbf{\bibinfo{volume}{119}},
  \bibinfo{pages}{216601} (\bibinfo{year}{2017}),
  \urlprefix\url{https://link.aps.org/doi/10.1103/PhysRevLett.119.216601}.

\bibitem[{\citenamefont{Davison et~al.}(2017)\citenamefont{Davison, Fu,
  Georges, Gu, Jensen, and Sachdev}}]{Gu2017b}
\bibinfo{author}{\bibfnamefont{R.~A.} \bibnamefont{Davison}},
  \bibinfo{author}{\bibfnamefont{W.}~\bibnamefont{Fu}},
  \bibinfo{author}{\bibfnamefont{A.}~\bibnamefont{Georges}},
  \bibinfo{author}{\bibfnamefont{Y.}~\bibnamefont{Gu}},
  \bibinfo{author}{\bibfnamefont{K.}~\bibnamefont{Jensen}}, \bibnamefont{and}
  \bibinfo{author}{\bibfnamefont{S.}~\bibnamefont{Sachdev}},
  \bibinfo{journal}{Phys. Rev. B} \textbf{\bibinfo{volume}{95}},
  \bibinfo{pages}{155131} (\bibinfo{year}{2017}),
  \urlprefix\url{https://link.aps.org/doi/10.1103/PhysRevB.95.155131}.

\bibitem[{\citenamefont{Patel et~al.}(2018)\citenamefont{Patel, McGreevy,
  Arovas, and Sachdev}}]{patel2017}
\bibinfo{author}{\bibfnamefont{A.~A.} \bibnamefont{Patel}},
  \bibinfo{author}{\bibfnamefont{J.}~\bibnamefont{McGreevy}},
  \bibinfo{author}{\bibfnamefont{D.~P.} \bibnamefont{Arovas}},
  \bibnamefont{and} \bibinfo{author}{\bibfnamefont{S.}~\bibnamefont{Sachdev}},
  \bibinfo{journal}{Phys. Rev. X} \textbf{\bibinfo{volume}{8}},
  \bibinfo{pages}{021049} (\bibinfo{year}{2018}),
  \urlprefix\url{https://link.aps.org/doi/10.1103/PhysRevX.8.021049}.

\bibitem[{\citenamefont{Chowdhury et~al.}(2018)\citenamefont{Chowdhury, Werman,
  Berg, and Senthil}}]{berg2018}
\bibinfo{author}{\bibfnamefont{D.}~\bibnamefont{Chowdhury}},
  \bibinfo{author}{\bibfnamefont{Y.}~\bibnamefont{Werman}},
  \bibinfo{author}{\bibfnamefont{E.}~\bibnamefont{Berg}}, \bibnamefont{and}
  \bibinfo{author}{\bibfnamefont{T.}~\bibnamefont{Senthil}},
  \bibinfo{journal}{Phys. Rev. X} \textbf{\bibinfo{volume}{8}},
  \bibinfo{pages}{031024} (\bibinfo{year}{2018}),
  \urlprefix\url{https://link.aps.org/doi/10.1103/PhysRevX.8.031024}.

\bibitem[{\citenamefont{Wu et~al.}(2018)\citenamefont{Wu, Chen, Jian, You, and
  Xu}}]{xu2018}
\bibinfo{author}{\bibfnamefont{X.}~\bibnamefont{Wu}},
  \bibinfo{author}{\bibfnamefont{X.}~\bibnamefont{Chen}},
  \bibinfo{author}{\bibfnamefont{C.-M.} \bibnamefont{Jian}},
  \bibinfo{author}{\bibfnamefont{Y.-Z.} \bibnamefont{You}}, \bibnamefont{and}
  \bibinfo{author}{\bibfnamefont{C.}~\bibnamefont{Xu}}, \bibinfo{journal}{Phys.
  Rev. B} \textbf{\bibinfo{volume}{98}}, \bibinfo{pages}{165117}
  (\bibinfo{year}{2018}),
  \urlprefix\url{https://link.aps.org/doi/10.1103/PhysRevB.98.165117}.

\bibitem[{\citenamefont{Gurau}(2011)}]{Gurau}
\bibinfo{author}{\bibfnamefont{R.}~\bibnamefont{Gurau}},
  \bibinfo{journal}{Commun. Math. Phys.} \textbf{\bibinfo{volume}{304}},
  \bibinfo{pages}{69} (\bibinfo{year}{2011}).

\bibitem[{\citenamefont{Mathur et~al.}(1998)\citenamefont{Mathur, Grosche,
  Julian, Walker, Freye, Haselwimmer, and Lonzarich}}]{rho0}
\bibinfo{author}{\bibfnamefont{N.~D.} \bibnamefont{Mathur}},
  \bibinfo{author}{\bibfnamefont{F.~M.} \bibnamefont{Grosche}},
  \bibinfo{author}{\bibfnamefont{S.~R.} \bibnamefont{Julian}},
  \bibinfo{author}{\bibfnamefont{I.~R.} \bibnamefont{Walker}},
  \bibinfo{author}{\bibfnamefont{D.~M.} \bibnamefont{Freye}},
  \bibinfo{author}{\bibfnamefont{R.~K.~W.} \bibnamefont{Haselwimmer}},
  \bibnamefont{and} \bibinfo{author}{\bibfnamefont{G.~G.}
  \bibnamefont{Lonzarich}}, \bibinfo{journal}{Nature}
  \textbf{\bibinfo{volume}{394}}, \bibinfo{pages}{39} (\bibinfo{year}{1998}).

\bibitem[{\citenamefont{Yuan et~al.}(2006)\citenamefont{Yuan, Grosche, Deppe,
  Sparn, Geibel, and Steglich}}]{cecusi}
\bibinfo{author}{\bibfnamefont{H.~Q.} \bibnamefont{Yuan}},
  \bibinfo{author}{\bibfnamefont{F.~M.} \bibnamefont{Grosche}},
  \bibinfo{author}{\bibfnamefont{M.}~\bibnamefont{Deppe}},
  \bibinfo{author}{\bibfnamefont{G.}~\bibnamefont{Sparn}},
  \bibinfo{author}{\bibfnamefont{C.}~\bibnamefont{Geibel}}, \bibnamefont{and}
  \bibinfo{author}{\bibfnamefont{F.}~\bibnamefont{Steglich}},
  \bibinfo{journal}{Phys. Rev. Lett.} \textbf{\bibinfo{volume}{96}},
  \bibinfo{pages}{047008} (\bibinfo{year}{2006}),
  \urlprefix\url{https://link.aps.org/doi/10.1103/PhysRevLett.96.047008}.

\bibitem[{\citenamefont{Tsujii et~al.}(2007)\citenamefont{Tsujii, Kitazawa,
  Aoyagi, Kimura, and Kido}}]{rho1}
\bibinfo{author}{\bibfnamefont{N.}~\bibnamefont{Tsujii}},
  \bibinfo{author}{\bibfnamefont{H.}~\bibnamefont{Kitazawa}},
  \bibinfo{author}{\bibfnamefont{T.}~\bibnamefont{Aoyagi}},
  \bibinfo{author}{\bibfnamefont{T.}~\bibnamefont{Kimura}}, \bibnamefont{and}
  \bibinfo{author}{\bibfnamefont{G.}~\bibnamefont{Kido}},
  \bibinfo{journal}{Journal of Magnetism and Magnetic Materials}
  \textbf{\bibinfo{volume}{310}}, \bibinfo{pages}{349 } (\bibinfo{year}{2007}),
  ISSN \bibinfo{issn}{0304-8853}, \bibinfo{note}{proceedings of the 17th
  International Conference on Magnetism},
  \urlprefix\url{http://www.sciencedirect.com/science/article/pii/S03048853060%
12248}.

\bibitem[{\citenamefont{Yuan and Steglich}(2007)}]{rho2}
\bibinfo{author}{\bibfnamefont{H.}~\bibnamefont{Yuan}} \bibnamefont{and}
  \bibinfo{author}{\bibfnamefont{F.}~\bibnamefont{Steglich}},
  \bibinfo{journal}{Physica C: Superconductivity and its Applications}
  \textbf{\bibinfo{volume}{460-462}}, \bibinfo{pages}{141 }
  (\bibinfo{year}{2007}), ISSN \bibinfo{issn}{0921-4534},
  \bibinfo{note}{proceedings of the 8th International Conference on Materials
  and Mechanisms of Superconductivity and High Temperature Superconductors},
  \urlprefix\url{http://www.sciencedirect.com/science/article/pii/S09214534070%
05242}.

\bibitem[{\citenamefont{Stockert and Steglich}(2011)}]{rho3}
\bibinfo{author}{\bibfnamefont{O.}~\bibnamefont{Stockert}} \bibnamefont{and}
  \bibinfo{author}{\bibfnamefont{F.}~\bibnamefont{Steglich}},
  \bibinfo{journal}{Annual Review of Condensed Matter Physics}
  \textbf{\bibinfo{volume}{2}}, \bibinfo{pages}{79} (\bibinfo{year}{2011}),
  \eprint{https://doi.org/10.1146/annurev-conmatphys-062910-140546},
  \urlprefix\url{https://doi.org/10.1146/annurev-conmatphys-062910-140546}.

\bibitem[{\citenamefont{L\"ohneysen
  et~al.}(2007{\natexlab{b}})\citenamefont{L\"ohneysen, Rosch, Vojta, and
  W\"olfle}}]{review}
\bibinfo{author}{\bibfnamefont{H.~v.} \bibnamefont{L\"ohneysen}},
  \bibinfo{author}{\bibfnamefont{A.}~\bibnamefont{Rosch}},
  \bibinfo{author}{\bibfnamefont{M.}~\bibnamefont{Vojta}}, \bibnamefont{and}
  \bibinfo{author}{\bibfnamefont{P.}~\bibnamefont{W\"olfle}},
  \bibinfo{journal}{Rev. Mod. Phys.} \textbf{\bibinfo{volume}{79}},
  \bibinfo{pages}{1015} (\bibinfo{year}{2007}{\natexlab{b}}),
  \urlprefix\url{https://link.aps.org/doi/10.1103/RevModPhys.79.1015}.

\bibitem[{\citenamefont{Georges et~al.}(2001)\citenamefont{Georges, Parcollet,
  and Sachdev}}]{sachdev2001}
\bibinfo{author}{\bibfnamefont{A.}~\bibnamefont{Georges}},
  \bibinfo{author}{\bibfnamefont{O.}~\bibnamefont{Parcollet}},
  \bibnamefont{and} \bibinfo{author}{\bibfnamefont{S.}~\bibnamefont{Sachdev}},
  \bibinfo{journal}{Phys. Rev. B} \textbf{\bibinfo{volume}{63}},
  \bibinfo{pages}{134406} (\bibinfo{year}{2001}),
  \urlprefix\url{https://link.aps.org/doi/10.1103/PhysRevB.63.134406}.

\bibitem[{\citenamefont{Dartois et~al.}(2017)\citenamefont{Dartois, Erbin, and
  Mondal}}]{nextorder}
\bibinfo{author}{\bibfnamefont{S.}~\bibnamefont{Dartois}},
  \bibinfo{author}{\bibfnamefont{H.}~\bibnamefont{Erbin}}, \bibnamefont{and}
  \bibinfo{author}{\bibfnamefont{S.}~\bibnamefont{Mondal}},
  \bibinfo{journal}{arXiv:1706.00412}  (\bibinfo{year}{2017}).

\end{thebibliography}

\end{document}